\documentclass[preprint,aps,amsmath,superscriptaddress,nofootinbib,tightenlines,showpacs]{revtex4}
\usepackage{amsmath}
\usepackage{amsfonts}
\usepackage{amssymb}
\usepackage{latexsym}
\usepackage{graphicx}
\usepackage{bm}
\usepackage{epsfig}
\usepackage[english]{babel}

\begin{document}

%\preprint{\vbox{ \hbox{}   \hbox{} }}

\title{Cosmological horizons as new examples of membrane paradigm}
\author{Tower Wang\footnote{Electronic address: twang@phy.ecnu.edu.cn}}
\affiliation{Shanghai Key Laboratory of Particle Physics and Cosmology,\\
Shanghai Jiao Tong University,\\
Shanghai 200240, China}
\affiliation{Department of Physics, East China Normal University,\\
Shanghai 200241, China\\ \vspace{0.2cm}}
\date{\today\\ \vspace{1cm}}
\begin{abstract}
In this paper we aim to provide new examples of the application and the generality of the membrane paradigm. The membrane paradigm is a formalism for studying the event horizon of black holes. After analyzing it with some technical details and realizing it in the Reissner-Nordstr\"om black hole, we apply the paradigm to cosmological horizons, firstly to the pure de Sitter horizon, and then to the trapping horizon of the Friedmann-Lema\^{i}tre-Robertson-Walker universe. In the latter case, the cosmological stretched horizon is oblique, thus the running of renormalization parameter is nonzero in the timelike direction and gives a correction to the membrane pressure. In this paradigm, the cosmological equations come from continuity equations of the membrane fluid and the bulk fluid respectively.
\end{abstract}


\maketitle

%\tighten

%%%%%%%%%%%%%%%%%%%%%%%%%%%%%%%%%%%%%%%%%%

%\tableofcontents
\section{Introduction}\label{sect-intro}
Black holes attract us not only with strong gravitational fields, but also with intriguing physical properties such as thermodynamics and hydrodynamics. The membrane paradigm is an appropriate formalism for studying the thermodynamic and hydrodynamic properties of black holes. On the event horizon of black hole, the gravitational equations resemble dynamical equations of a 2-dimensional viscous fluid, including the energy conservation law and the Navier-Stokes equation\cite{Damour79,Damour82,Damour:2008ji}. However, it is unpractical for a distant observer to study this fluid-like behavior directly on the event horizon, because the event horizon's generators are not timelike but null. The membrane paradigm for gravitational fields around a black hole was constructed three decades ago \cite{Price:1986yy,Thorne86} on the stretched horizon, a timelike hypersurface located slightly outside the event horizon. Fortunately, the dynamical equations of the event horizon are well approximated by those of the stretched horizon. Thus from outside it is more convenient to probe the fluid-like behavior of black holes on the stretched horizon. In particular, up to a renormalization factor, from the membrane on the stretched horizon, we can read the fluid quantities such as density, pressure, shear and expansion. The renormalization parameter is presumed to tend to zero as the stretched horizon approaches the event horizon.

The membrane paradigm were based on equations of motion and partially on various intuitive physical arguments until one and a half decades ago, when Parikh \emph{et al} \cite{Parikh:1997ma} systematically derived the above results from an action with a surface term on the stretched horizon. The Raychaudhuri equation, or the energy conservation law of membrane fluid, was obtained as the timelike component of the contracted Gauss-Codazzi equations, while the transverse components on the membrane give rise to the Navier-Stokes equation. This is a more rigorous and elegant approach to the black hole membrane paradigm. Taking this approach, the paradigm was studied in some other theories of gravity \cite{Chatterjee:2010gp, Jacobson:2011dz,Kolekar:2011gg}. Interesting related work can be also found in \cite{Ashtekar:2000hw,Booth:2001gx,Padmanabhan:2010rp,Jaramillo:2013rda} as a very incomplete list.

Usually when a paradigm becomes more rigorous in formulation, its implicit assumptions and restrictions will get less elusive. In the present paper, we will show this indeed happens to the black hole membrane paradigm, and we will apply the paradigm to cosmological horizons. Following the approach of reference \cite{Parikh:1997ma}, we will present a brief review of the membrane paradigm in section \ref{sect-rev} and Appendix \ref{app-der}, and investigate the membrane paradigm in a concrete example---the Reissner-Nordstr\"om (RN) black hole in section \ref{sect-RN}. After that, we will express new examples of the paradigm through cosmological horizons: the de Sitter (dS) horizon in section \ref{sect-dS} and the trapping horizon of the Friedmann-Lema\^{i}tre-Robertson-Walker (FLRW) universe \cite{Hayward:1993ph,Cai:2006rs} in section \ref{sect-FLRW}. In section \ref{sect-Friedmann}, the Friedmann equation emerges naturally from the membrane paradigm. We demonstrate in section \ref{sect-rev} and illustrate in section \ref{sect-FLRW} that the accurate value of pressure on the event/trapping horizon may deviate from the renormalized pressure on the stretched horizon, by receiving a correction from running of the renormalization parameter. In section \ref{sect-FLRW}, for the FLRW universe, the renormalization parameter is time-dependent, and hence its derivative is not zero along the time-like direction. That is why the paradigm therein is called an oblique membrane paradigm. In section \ref{sect-stand}, we will ponder on a tentative way towards the standard membrane paradigm that is not oblique. The paper is concluded in section \ref{sect-con}.

A remark is needed here. By the name ``oblique membrane paradigm'', we do not mean development of a new paradigm, but providing examples of the existing membrane paradigm to the cosmological horizons and the specific non-stationary horizons. In other words, our best wish is to present the application of the membrane paradigm to the cosmological horizon, and provide an explicit example (FLRW) where the trapping horizon is time dependent. In fact, it is very clear from the original derivation of the black-hole membrane paradigm or the derivation a la Parikh-Wilczek \cite{Parikh:1997ma} that the so called pressure of the fluid depends upon the choice of the generators. It agrees with the surface gravity only in the case that the horizon is stationary. Viewed by top scientists of black hole, this paper may not contain any new scientific result, but we hope it would be of some use for researchers on cosmology. We wish there will be connections of the membrane paradigm to observational cosmology, such as cosmic inflation and accelerated expansion.

\section{Review of membrane paradigm}\label{sect-rev}
In this section, we will briefly review the membrane paradigm following the approach of \cite{Parikh:1997ma}, with an emphasis on some technical details necessary for our analysis but hidden in the literature.

\subsection{Geometric setup}\label{subsect-geom}
It is customary to begin with the geometry of spacetime and our convention of notations. For the simplest example, we will work with a spherically symmetric event or trapping null horizon in $(3+1)$-dimensional spacetime and mainly focus on the Einstein gravity. With appropriate extensions our calculation is potentially applicable to other cases.

The event horizon is a 3-dimensional null hypersurface with a null geodesic generator $l^{a}$. At the event horizon, the surface gravity $g_{\mathcal{H}}$ can be determined by
\begin{equation}\label{surfg}
l^{b}\nabla_{b}l^{a}=g_{\mathcal{H}}l^{a}.
\end{equation}
Besides this one, there are several other ways to define the surface gravity \cite{Jacobson:1993pf}. For Killing horizons such as the event horizon of the RN black hole, all definitions yield the same result. But for non-Killing horizons, \emph{e.g.} the trapping horizon of the FLRW universe, different definitions disagree with each other \cite{Jacobson:1993pf,Cropp:2013zxi}. In this paper, we will choose equation \eqref{surfg} as the definition of surface gravity at non-Killing null horizons. For the spherically symmetric case, references \cite{Fodor:1996rf,Nielsen:2007ac} are in favor of this choice.

In the very neighborhood of event/trapping horizon, suppose there is a 3-dimensional timelike hypersurface, namely stretched horizon, which is generated by the timelike congruence $u^{a}$ and possesses a spacelike normal vector $n^{a}$. The vectors $u^{a}$ and $n^{a}$ are orthogonal to each other
\begin{equation}\label{orth}
n_{a}u^{a}=0
\end{equation}
and normalized to unity
\begin{equation}\label{norm}
n_{a}n^{a}=1,~~~~u_{a}u^{a}=-1.
\end{equation}
%\begin{eqnarray}
%\label{orth}&&n_{a}u^{a}=0,\\
%\label{norm}&&n_{a}n^{a}=1,~~~~u_{a}u^{a}=-1.
%\end{eqnarray}
In terms of $u^{a}$ and $n^{a}$, we can make a $2+1+1$ split of spacetime, which leads to a 3-dimensional metric
\begin{equation}\label{3metric}
h_{ab}=g_{ab}-n_{a}n_{b}
\end{equation}
on the stretched horizon and a 2-dimensional metric
\begin{equation}\label{2metric}
\gamma_{ab}=h_{ab}+u_{a}u_{b}
\end{equation}
on its spacelike cross section normal to $u^{a}$. Then it is straightforward to express the stretched horizon's extrinsic curvature
\begin{equation}\label{3k}
%K_{ab}=h_{a}^{c}h_{b}^{d}\nabla_{d}n_{c}=h^{d}_{b}\nabla_{d}n_{a}.
K_{ab}=h^{d}_{b}\nabla_{d}n_{a}.
\end{equation}

In the membrane paradigm, the stretched horizon plays the role of an auxiliary hypersurface tracking the event horizon. To this end, a parameter $\alpha$ is introduced, which can be taken as a regulator. When the stretched horizon tends to the true horizon, it is required that $\alpha\rightarrow0$, $\alpha u^{a}\rightarrow l^{a}$ and $\alpha n^{a}\rightarrow l^{a}$. This requirement is in accord with the fact that the null generator $l^{a}$ is both normal and tangential to the event horizon, namely $l_{a}l^{a}=\gamma_{ab}l^{a}=0$. Consequently in this limit we expect
\begin{eqnarray}
\label{klimg}\alpha u^{a}u^{b}K_{ab}=u^{a}u^{b}\nabla_{b}(\alpha n_{a})&\rightarrow&-g_{\mathcal{H}},\\
\label{klimk}\alpha\gamma_{A}^{a}\gamma_{B}^{b}K_{ab}=\gamma_{A}^{a}\gamma_{B}^{b}\nabla_{b}(\alpha n_{a})&\rightarrow&k_{AB},\\
\label{dulimk}\gamma_{A}^{a}\gamma_{B}^{b}\nabla_{b}(\alpha u_{a})&\rightarrow&k_{AB},
\end{eqnarray}
where $k_{AB}$ is the extrinsic curvature of the 2-dimensional spacelike section of the event horizon,
\begin{equation}\label{2k}
k_{AB}=\gamma_{A}^{a}\gamma_{B}^{b}\nabla_{b}l_{a}=\frac{1}{2}\mathcal{L}_{l}\gamma_{AB}
\end{equation}
with $\mathcal{L}_{l}$ the Lie derivative in the direction of $l^{a}$. It will be useful to decompose $k_{AB}$ into a traceless part and a trace,
\begin{equation}\label{2kdec}
k_{AB}=\sigma_{\mathcal{H}AB}+\frac{1}{2}\theta_{\mathcal{H}}\gamma_{AB},
\end{equation}
where $\sigma_{\mathcal{H}AB}$ is the shear and $\theta_{\mathcal{H}}$ is the expansion of the world lines of nearby horizon surface elements.

To study the dynamics of null horizons, it is useful to introduce the H\'a\'{\j}i\v{c}ek field $\Omega_{\mathcal{H}A}$ in terms of the outgoing null vector $l^{a}$ and an ingoing null vector \cite{Price:1986yy,Gourgoulhon:2005ng}. Translated into $u^{a}$ and $n^{a}$ in our cases, it reads
\begin{equation}\label{Hajek}
\Omega_{\mathcal{H}A}=\frac{1}{2}\gamma_{A}^{a}(n^{b}\nabla_{b}u_{a}-u^{b}\nabla_{b}n_{a}).
\end{equation}
In the near event horizon limit we expect
\begin{equation}\label{klimHaj}
\gamma_{A}^{a}u^{b}K_{ab}=\gamma_{A}^{a}u^{b}\nabla_{b}n_{a}\rightarrow-\Omega_{\mathcal{H}A}
\end{equation}
or equivalently
\begin{equation}\label{klimdu}
\gamma_{A}^{a}u^{b}K_{ab}\rightarrow-\gamma_{A}^{a}n^{b}\nabla_{b}u_{a}.
\end{equation}
Without loss of generality, one can parameterize the the spacelike normal vector as $n_{a}=N\nabla_{a}\lambda$ and the timelike generator as $u_{a}=-Uh_{a}^{b}\nabla_{b}\tau$, where $\lambda$ and $\tau$ are affine or non-affine parameters. In appendix \ref{app-der}, we have proven that
\begin{eqnarray}
\nonumber n^{a}\nabla_{a}n^{b}&=&-\gamma^{ab}\nabla_{a}\ln N+u^{b}u^{a}\nabla_{a}\ln N,\\
u^{a}\nabla_{a}u^{b}&=&\gamma^{ab}\nabla_{a}\ln U-n^{b}K_{ac}u^{a}u^{c}.
\end{eqnarray}
For the membrane paradigm of RN black holes and the paradigm of cosmological horizons, it can be confirmed that $\gamma^{ab}\nabla_{a}\ln N=\gamma^{ab}\nabla_{a}\ln U=0$. Therefore, throughout this paper, it is safe to make the ansatz
\begin{equation}\label{ansatz}
n^{a}\nabla_{a}n^{b}=u^{b}u^{a}\nabla_{a}\ln N,~~~~u^{a}\nabla_{a}u^{b}=-n^{b}K_{ac}u^{a}u^{c}.
\end{equation}
Remembering that $(u^{b}+n^{b})\nabla_{b}(u_{a}+n_{a})\propto u_{a}+n_{a}$ and thus $\gamma_{A}^{a}(u^{b}+n^{b})\nabla_{b}(u_{a}+n_{a})=0$, we see limit \eqref{klimHaj} can be always satisfied exactly under this ansatz.

\subsection{Fluid equations from membrane dynamics}\label{subsect-flueq}
In reference \cite{Parikh:1997ma}, it has been verified that for Einstein gravity, the membrane stress tensor on the stretched horizon takes the form\footnote{Derived in reference \cite{Parikh:1997ma} for stretched horizon outside the true horizon, the signature of $t_{\mathcal{S}ab}$ is unchanged for stretched horizon inside the true horizon, because we will then deal with $\delta S_{in}-\delta S_{surf}=0$ and an inward-pointing normal vector $n^{a}$.}
\begin{equation}\label{tK}
t_{\mathcal{S}ab}=\frac{1}{8\pi G}\left(Kh_{ab}-K_{ab}\right).
\end{equation}
For our purpose, it is convenient to reverse this equation to
\begin{equation}\label{Kt}
K_{ab}=8\pi G\left(\frac{1}{2}t_{\mathcal{S}d}^{d}h_{ab}-t_{\mathcal{S}ab}\right).
\end{equation}
Interestingly, the above stress tensor can be decomposed to a form like a viscous fluid \cite{Jacobson:2011dz},
\begin{equation}
t_{\mathcal{S}}^{ab}=\rho_{\mathcal{S}}u^{a}u^{b}+\gamma_{A}^{a}\gamma_{B}^{b}(p_{\mathcal{S}}\gamma^{AB}-2\eta_{\mathcal{S}}\sigma_{\mathcal{S}}^{AB}-\zeta_{\mathcal{S}}\theta_{\mathcal{S}}\gamma^{AB})+\pi_{\mathcal{S}}^{A}(\gamma_{A}^{a}u^{b}+\gamma_{A}^{b}u^{a}).
\end{equation}
Some of the fluid quantities defined above, including density $\rho_{\mathcal{S}}$, pressure $p_{\mathcal{S}}$, shear $\sigma_{\mathcal{S}}^{AB}$ and expansion $\theta_{\mathcal{S}}$ on the stretched horizon, get divergent in the near event horizon limit $\alpha\rightarrow0$. We can renormalize them with the regulator $\alpha$,
\begin{equation}
\rho_{\mathcal{S}}=\frac{1}{\alpha}\rho_{\mathcal{H}},~~~~p_{\mathcal{S}}=\frac{1}{\alpha}p_{\mathcal{H}},~~~~\sigma_{\mathcal{S}}^{AB}=\frac{1}{\alpha}\sigma_{\mathcal{H}}^{AB},~~~~\theta_{\mathcal{S}}=\frac{1}{\alpha}\theta_{\mathcal{H}},
\end{equation}
%\begin{eqnarray}
%\nonumber&&\rho_{\mathcal{S}}=\frac{1}{\alpha}\rho_{\mathcal{H}},~~~~p_{\mathcal{S}}=\frac{1}{\alpha}p_{\mathcal{H}},\\
%&&\sigma_{\mathcal{S}}^{AB}=\frac{1}{\alpha}\sigma_{\mathcal{H}}^{AB},~~~~\theta_{\mathcal{S}}=\frac{1}{\alpha}\theta_{\mathcal{H}},
%\end{eqnarray}
keep $\pi_{\mathcal{S}}^{A}=\pi_{\mathcal{H}}^{A}$, $\eta_{\mathcal{S}}=\eta_{\mathcal{H}}$, $\zeta_{\mathcal{S}}=\zeta_{\mathcal{H}}$, and rewrite the membrane stress tensor as
\begin{equation}\label{mem-stress}
t_{\mathcal{S}}^{ab}=\frac{1}{\alpha}\rho_{\mathcal{H}}u^{a}u^{b}+\frac{1}{\alpha}\gamma_{A}^{a}\gamma_{B}^{b}(p_{\mathcal{H}}\gamma^{AB}-2\eta_{\mathcal{H}}\sigma_{\mathcal{H}}^{AB}-\zeta_{\mathcal{H}}\theta_{\mathcal{H}}\gamma^{AB})+\pi_{\mathcal{H}}^{A}(\gamma_{A}^{a}u^{b}+\gamma_{A}^{b}u^{a}).
\end{equation}

Substituting \eqref{mem-stress} into \eqref{Kt}, we can demonstrate that
\begin{eqnarray}
\nonumber\alpha K_{ab}\gamma_{A}^{a}\gamma_{B}^{b}&=&8\pi G\left(-\frac{1}{2}\rho_{\mathcal{H}}\gamma_{AB}+2\eta_{\mathcal{H}}\sigma_{\mathcal{H}AB}\right),\\
\nonumber\alpha K_{ab}u^{a}u^{b}&=&8\pi G\left(\zeta_{\mathcal{H}}\theta_{\mathcal{H}}-\frac{1}{2}\rho_{\mathcal{H}}-p_{\mathcal{H}}\right),\\
K_{ab}\gamma_{A}^{a}u^{b}&=&8\pi G\pi_{\mathcal{H}A}.
\end{eqnarray}
Thus the limit \eqref{klimk} can be fulfilled by
\begin{equation}\label{klimcond1}
\eta_{\mathcal{H}}=\frac{1}{16\pi G},~~~~\rho_{\mathcal{H}}=-\frac{\theta_{\mathcal{H}}}{8\pi G},
\end{equation}
while the limit \eqref{klimg} can be achieved if in addition
\begin{equation}\label{klimcond2}
\zeta_{\mathcal{H}}=-\frac{1}{16\pi G},~~~~p_{\mathcal{H}}=\frac{g_{\mathcal{H}}}{8\pi G}.
\end{equation}
However, as will be shown soon in this section and later in section \ref{sect-FLRW}, the second equation of \eqref{klimcond2} is sometimes violated, see equations \eqref{Raylimcond1} and \eqref{Raylimcond2}. In that case, the limit \eqref{klimg} is replaced by \eqref{klimgmod}, and the accurate value of pressure on the event/trapping horizon is $g_{\mathcal{H}}/(8\pi G)$ but not $p_{\mathcal{H}}$.

In both Einstein and $f(R)$ theories of gravity \cite{Parikh:1997ma,Chatterjee:2010gp}, the contracted Gauss-Codazzi relation \cite{Gourgoulhon:2007ue} can be put into the form
\begin{equation}\label{Gauss-Codazzi}
t_{\mathcal{S}|b}^{ab}=-h_{c}^{a}T^{cd}n_{d},
\end{equation}
where $|b$ is the 3-covariant derivative with respect to the metric $h_{ab}$, and thus $t_{\mathcal{S}|b}^{ab}=h_{d}^{a}h_{b}^{c}\nabla_{c}t_{\mathcal{S}}^{db}$. Later on we will use the notation $||b$ for the 2-covariant derivative with respect to the metric $\gamma_{ab}$, and similarly $\sigma_{\mathcal{H}||b}^{ab}=\gamma_{d}^{a}\gamma_{b}^{c}\nabla_{c}\sigma_{\mathcal{H}}^{db}$. It is trivial to check $h_{ab|c}=\gamma_{ab||c}=0$.

Projecting \eqref{Gauss-Codazzi} to the timelike direction $u^{a}$, we have derived in Appendix \ref{app-der} that
\begin{eqnarray}
\nonumber-\frac{1}{\alpha^2}T_{b}^{a}l_{a}l^{b}&=&-\frac{1}{\alpha^2}(p_{\mathcal{H}}\gamma^{AB}-2\eta_{\mathcal{H}}\sigma_{\mathcal{H}}^{AB}-\zeta_{\mathcal{H}}\theta_{\mathcal{H}}\gamma^{AB})k_{AB}-\nabla_{b}\pi_{\mathcal{H}}^{b}\\
&&-\frac{1}{\alpha^2}l^{b}\nabla_{b}\rho_{\mathcal{H}}+\frac{1}{\alpha^2}\rho_{\mathcal{H}}u^{b}\nabla_{b}\alpha-\frac{1}{\alpha^2}\rho_{\mathcal{H}}\gamma_{b}^{a}k_{a}^{b}
\end{eqnarray}
when the stretched horizon gets close to the true horizon. The expression of $k_{AB}$ was given in \eqref{2kdec}, which helps to write down
%\begin{equation}
%-\frac{1}{\alpha^2}T_{b}^{a}l_{a}l^{b}+T_{ab}\left(\frac{1}{\alpha^2}l^{a}l^{b}-n^{a}u^{b}\right)=-\frac{1}{\alpha^2}(p_{\mathcal{H}}\theta_{\mathcal{H}}-2\eta_{\mathcal{H}}\sigma_{\mathcal{H}AB}\sigma_{\mathcal{H}}^{AB}-\zeta_{\mathcal{H}}\theta_{\mathcal{H}}^2)-\nabla_{b}\pi_{\mathcal{H}}^{b}-\frac{1}{\alpha^2}\mathcal{L}_{l}\rho_{\mathcal{H}}+\frac{1}{\alpha^2}\rho_{\mathcal{H}}\mathcal{L}_{l}\ln\alpha-\frac{1}{\alpha^2}\rho_{\mathcal{H}}\theta_{\mathcal{H}}.
%\end{equation}
\begin{eqnarray}\label{Raystr}
\nonumber-\frac{1}{\alpha^2}T_{b}^{a}l_{a}l^{b}&=&-\frac{1}{\alpha^2}(p_{\mathcal{H}}\theta_{\mathcal{H}}-2\eta_{\mathcal{H}}\sigma_{\mathcal{H}AB}\sigma_{\mathcal{H}}^{AB}-\zeta_{\mathcal{H}}\theta_{\mathcal{H}}^2)-\nabla_{b}\pi_{\mathcal{H}}^{b}\\
&&-\frac{1}{\alpha^2}\mathcal{L}_{l}\rho_{\mathcal{H}}+\frac{1}{\alpha^2}\rho_{\mathcal{H}}\mathcal{L}_{u}\alpha-\frac{1}{\alpha^2}\rho_{\mathcal{H}}\theta_{\mathcal{H}}.
\end{eqnarray}
On the other hand, the Raychaudhuri equation of a null geodesic congruence \cite{Gourgoulhon:2005ng}
\begin{equation}\label{Ray1}
l^{a}\nabla_{a}\theta_{\mathcal{H}}-g_{\mathcal{H}}\theta_{\mathcal{H}}+\frac{1}{2}\theta_{\mathcal{H}}^2+\sigma_{\mathcal{H}AB}\sigma_{\mathcal{H}}^{AB}+R_{b}^{a}l_{a}l^{b}=0
\end{equation}
can be written in a form similar to the energy conservation law of fluid
\begin{equation}\label{Ray2}
\mathcal{L}_{l}\rho_{\mathcal{H}}+\theta_{\mathcal{H}}\rho_{\mathcal{H}}=-\tilde{p}_{\mathcal{H}}\theta_{\mathcal{H}}+2\eta_{\mathcal{H}}\sigma_{\mathcal{H}AB}\sigma_{\mathcal{H}}^{AB}+\zeta_{\mathcal{H}}\theta_{\mathcal{H}}^2+T_{b}^{a}l_{a}l^{b}
\end{equation}
for the Einstein gravity. Here $\eta_{\mathcal{H}}$, $\zeta_{\mathcal{H}}$, $\rho_{\mathcal{H}}$ and $\tilde{p}_{\mathcal{H}}$ are given by equations \eqref{klimcond1}, \eqref{klimcond2} and \eqref{Raylimcond1}. In reference \cite{Gourgoulhon:2005ng}, equation \eqref{Ray1} above and \eqref{NS1} below were derived from a null analog of the contracted Gauss-Codazzi relation, \emph{i.e.} directly on the null true horizon, rather than from \eqref{Gauss-Codazzi} on the timelike stretched horizon.

In equation \eqref{Ray2} we have deliberately put a tilde above $p_{\mathcal{H}}$. Naively one may obtain \eqref{Ray2} from equation \eqref{Raystr} by equating $p_{\mathcal{H}}$ to $\tilde{p}_{\mathcal{H}}$ and neglecting the redundant terms. As we send $\alpha$ to zero, the $\nabla_{b}\pi_{\mathcal{S}}^{b}$ term is negligible as expected, but the $\rho_{\mathcal{H}}\mathcal{L}_{u}\alpha$ term does not always vanish. Actually, as we will see in section \ref{sect-FLRW}, in the case of the FLRW universe, $\alpha$ is time-dependent and thus the $\rho_{\mathcal{H}}\mathcal{L}_{u}\alpha$ term is nonvanishing. When this happens, the stretched horizon will be dubbed an oblique membrane.

The key point is, as we have mentioned, that when the membrane is oblique, the second equation of \eqref{klimcond2} is violated, or more exactly
\begin{equation}\label{Raylimcond1}
\tilde{p}_{\mathcal{H}}=\frac{g_{\mathcal{H}}}{8\pi G}\neq p_{\mathcal{H}}.
\end{equation}
To recover equation \eqref{Ray2} from \eqref{Raystr}, the effective pressure $\tilde{p}_{\mathcal{H}}$ on the event/trapping horizon should be related to $p_{\mathcal{H}}$ via
\begin{equation}
-\frac{1}{\alpha^2}\tilde{p}_{\mathcal{H}}\theta_{\mathcal{H}}=-\frac{1}{\alpha^2}p_{\mathcal{H}}\theta_{\mathcal{H}}+\frac{1}{\alpha^2}\rho_{\mathcal{H}}\mathcal{L}_{u}\alpha,
\end{equation}
or equivalently
\begin{equation}\label{Raylimcond2}
\tilde{p}_{\mathcal{H}}=p_{\mathcal{H}}+\frac{1}{8\pi G}\mathcal{L}_{u}\alpha
\end{equation}
thanks to the second equation of \eqref{klimcond1}. This relation is deduced by comparing \eqref{Raystr} with \eqref{Ray2}. It will be further confirmed by examples in this paper. Note that \eqref{klimcond1} and the first equation of \eqref{klimcond2} are not harmed, so we still have
\begin{equation}\label{Raylimcond3}
\alpha K_{ab}u^{a}u^{b}=-8\pi Gp_{\mathcal{H}},
\end{equation}
but the limit \eqref{klimg} is replaced by
\begin{equation}\label{klimgmod}
\alpha u^{a}u^{b}K_{ab}-\mathcal{L}_{u}\alpha\rightarrow-g_{\mathcal{H}}.
\end{equation}

Starting with equations \eqref{mem-stress} and \eqref{Gauss-Codazzi}, in Appendix \ref{app-der} we have proven
\begin{equation}\label{NSstr}
\gamma_{A}^{e}\mathcal{L}_{l}\pi_{\mathcal{H}e}+\pi_{\mathcal{H}A}\theta_{\mathcal{H}}=-p_{\mathcal{H}||A}+2(\eta_{\mathcal{H}}\sigma_{\mathcal{H}A}^{B})_{||B}+(\zeta_{\mathcal{H}}\theta_{\mathcal{H}})_{||A}-T_{a}^{c}l_{c}\gamma_{A}^{a}.
\end{equation}
On the other hand, the H\'a\'{\j}i\v{c}ek equation \cite{Gourgoulhon:2005ng}
\begin{equation}\label{NS1}
R_{a}^{c}l_{c}\gamma_{A}^{a}=\gamma_{A}^{e}\mathcal{L}_{l}\Omega_{\mathcal{H}e}+\theta_{\mathcal{H}}\Omega_{\mathcal{H}A}-\left(g_{\mathcal{H}}+\frac{1}{2}\theta_{\mathcal{H}}\right)_{||A}+\sigma_{\mathcal{H}A||B}^{B}
\end{equation}
can be written in a form similar to the Navier-Stokes equation of fluid
\begin{equation}\label{NS2}
\gamma_{A}^{e}\mathcal{L}_{l}\pi_{\mathcal{H}e}+\pi_{\mathcal{H}A}\theta_{\mathcal{H}}=-\tilde{p}_{\mathcal{H}||A}+2(\eta_{\mathcal{H}}\sigma_{\mathcal{H}A}^{B})_{||B}+(\zeta_{\mathcal{H}}\theta_{\mathcal{H}})_{||A}-T_{a}^{c}l_{c}\gamma_{A}^{a}
\end{equation}
for the Einstein gravity. Here again $\eta_{\mathcal{H}}$, $\zeta_{\mathcal{H}}$, $\rho_{\mathcal{H}}$ and $\tilde{p}_{\mathcal{H}}$ are given by equations \eqref{klimcond1}, \eqref{klimcond2} and \eqref{Raylimcond1}, and
\begin{equation}
\pi_{\mathcal{H}}^{a}=-\frac{1}{8\pi G}\Omega_{\mathcal{H}}^{a}
\end{equation}
which is based on the limit \eqref{klimHaj} or \eqref{klimdu}.

At first glance, equation \eqref{NS2} is in discrepancy with \eqref{NSstr} in case $\tilde{p}_{\mathcal{H}}\neq p_{\mathcal{H}}$. The discrepancy can be resolved if
\begin{equation}\label{NSlimcond}
p_{\mathcal{H}||A}=\tilde{p}_{\mathcal{H}||A},
\end{equation}
which indeed holds for spherically symmetric horizons at the least. For horizons without spherical symmetry, we do not have a general proof for this equality hitherto, but we have checked that equality \eqref{NSlimcond} continues to hold for Kerr-Newman black holes.

A generalized form of equations \eqref{Ray2} and \eqref{NS2} for non-null hypersurfaces can be found in \cite{Gourgoulhon:2005ch,Gourgoulhon:2006uc}, although our present paper will be restricted to null horizons.

\section{Memberane paradigm for RN black hole}\label{sect-RN}
As an illuminating example, we will work out the quantities and equations of the membrane paradigm for the RN black hole. The RN solution in $3+1$ dimensions has the form
\begin{equation}\label{RN}
ds^2=-f(r)dt^2+\frac{1}{f(r)}dr^2+r^2d\Omega^2
\end{equation}
with $f(r)=1-2Mr^{-1}+Q^2r^{-2}$ and
$d\Omega^2=d\vartheta^2+\sin^2\vartheta d\varphi^2$. In the membrane paradigm \cite{Price:1986yy,Parikh:1997ma}, the stretched horizon of RN black holes has an outward-pointing spacelike unit normal $n^{a}\partial_{a}=f^{1/2}\partial_{r}$ and a future-directed timelike generator $u^{a}\partial_{a}=f^{-1/2}\partial_{t}$. Their dual form can be expressed as $n_{a}=N\nabla_{a}r$, $u_{a}=-Uh_{a}^{b}\nabla_{b}t$ with $N=f^{-1/2}$ and $U=f^{1/2}$, which agree well with ansatz \eqref{ansatz}. The bulk stress tensor is
\begin{equation}
8\pi GT_{ab}dx^{a}dx^{b}=\frac{Q^2f}{r^4}dt^2-\frac{Q^2}{r^4f}dr^2+\frac{Q^2}{r^2}d\Omega^2,
\end{equation}
which implies
\begin{equation}
T_{ab}n^{a}u^{b}=0.
\end{equation}
By definition of the 3-dimensional extrinsic curvature \eqref{3k}, we find the nonvanishing components
\begin{equation}
K_{tt}=\frac{f^{1/2}(Q^2-Mr)}{r^3},~~~~K_{\vartheta\vartheta}=rf^{1/2},~~~~K_{\varphi\varphi}=rf^{1/2}\sin^2\vartheta.
\end{equation}
Mapped to the temporal or transverse directions,
\begin{equation}\label{RN3k}
\alpha u^{a}u^{b}K_{ab}=\frac{\alpha(Q^2-Mr)}{r^3f^{1/2}},~~~~\alpha\gamma_{A}^{a}\gamma_{B}^{b}K_{ab}=\frac{\alpha f^{1/2}}{r}\gamma_{AB},~~~~\gamma_{A}^{a}u^{b}K_{ab}=0.
\end{equation}
One may also check that
\begin{equation}\label{RNdu}
\gamma_{A}^{a}\gamma_{B}^{b}\nabla_{b}(\alpha u_{a})=0,~~~~\gamma_{A}^{a}n^{b}\nabla_{b}u_{a}=0.
\end{equation}

In the Eddington-Finkelstein (EF) coordinates $dv=dt+f^{-1}dr$, the RN metric takes the form
\begin{equation}\label{RN-EF}
ds^2=-fdv^2+2dvdr+r^2d\Omega^2,
\end{equation}
while the unit normal and the timelike generator become
\begin{eqnarray}\label{RN-EFnu}
\nonumber n^{a}\partial_{a}&=&f^{-1/2}\partial_{v}+f^{1/2}\partial_{r},\\
u^{a}\partial_{a}&=&f^{-1/2}\partial_{v}.
\end{eqnarray}

The Kruskal-Szekeres (KS) coordinates $x$ and $y$ for the RN black hole are defined by equations
\begin{eqnarray}
\nonumber xy&=&-e^{2g_{+}r}\left(\frac{r-r_{+}}{r_{+}}\right)\left(\frac{r-r_{-}}{r_{-}}\right)^{g_{+}/g_{-}},\\
\frac{x}{y}&=&-e^{2g_{+}t}
\end{eqnarray}
with $r_{\pm}=M\pm\sqrt{M^2-Q^2}$ and $g_{\pm}=(r_{\pm}-r_{\mp})/(2r_{\pm}^2)$. In the KS coordinates, the metric covering the region $r\geq r_{-}$ is of the form
\begin{equation}\label{RN-KS}
ds^2=-\frac{r_{+}r_{-}}{g_{+}^2}\frac{e^{-2g_{+}r}}{r^2}\left(\frac{r_{-}}{r-r_{-}}\right)^{\frac{g_{+}}{g_{-}}-1}dxdy+r^2d\Omega^2,
\end{equation}
while the unit normal and the timelike generator are transformed to
\begin{eqnarray}\label{RN-KSnu}
\nonumber n^{a}\partial_{a}&=&f^{-1/2}g_{+}(x\partial_{x}+y\partial_{y}),\\
u^{a}\partial_{a}&=&f^{-1/2}g_{+}(x\partial_{x}-y\partial_{y}).
\end{eqnarray}
The future event horizon is the $y=0$ hypersurface.

The event horizon of RN black hole is a 3-dimensional null Killing horizon, whose generator satisfies both the null condition and the Killing equation locally. But the full $(3+1)$-dimensional spacetime does not own a null Killing field globally. As a consequence, outside the hypersurface we should break either the null condition or the Killing equation. We will study the RN black hole horizon with a globally Killing generator in section \ref{subsect-RNKil} and a globally null generator in section \ref{subsect-RNnul}. Restricted to the event horizon, each of them satisfies both the Killing equation and the null condition.

\subsection{Killing vector as horizon's generator}\label{subsect-RNKil}
For a stationary black hole there is a Killing vector
\begin{equation}
l^{a}\partial_{a}=\partial_{t}=f^{1/2}u^{a}\partial_{a}.
\end{equation}
This vector is null on the event horizon but non-null away from it. If we use this vector to extend the horizon's generator to the full spacetime, it is trivial to read off $\alpha=f^{1/2}$ (recall that $\alpha u^{a}\rightarrow l^{a}$ in the membrane paradigm) and
\begin{equation}
T_{ab}l^{a}l^{b}=\frac{Q^2}{8\pi Gr^4}f.
\end{equation}
Utilizing definition \eqref{2k}, we find the 2-dimensional extrinsic curvature vanishes
\begin{equation}\label{RNKil-2k}
k_{AB}=0.
\end{equation}
The surface gravity cannot be worked out directly with formula \eqref{surfg} in metric \eqref{RN}, but in the EF coordinates \eqref{RN-EF} it gives
\begin{equation}\label{RNKil-g}
g_{\mathcal{H}}=\frac{Mr-Q^2}{r^3}.
\end{equation}

With equations \eqref{RN3k}, \eqref{RNdu}, \eqref{RNKil-2k} and \eqref{RNKil-g} in hand, it is easy to check relations \eqref{klimg}, \eqref{klimk}, \eqref{dulimk} and \eqref{klimdu} in the limit $\alpha\rightarrow0$. In the present situation we have $u^{b}\nabla_{b}\alpha=f^{-1/2}\partial_{t}f^{1/2}=0$, hence the effective pressure on the event horizon coincides with the renormalized pressure on the stretched horizon, and equation \eqref{Raylimcond2} is simplified to $\tilde{p}_{\mathcal{H}}=p_{\mathcal{H}}=(Mr-Q^2)/(8\pi Gr^3)$, satisfying condition \eqref{NSlimcond} trivially.

The EF coordinates is best-suited for producing the limit $\alpha n^{a}\rightarrow l^{a}$. Leaving this to the reader for dessert, we now move on to another form of horizon's generator.

\subsection{Null vector as horizon's generator}\label{subsect-RNnul}
The RN solution has two principal null vectors. We can extend the event horizon's generator with the outgoing one of them,
\begin{equation}\label{lRN}
l^{a}\partial_{a}=\frac{1}{2}\partial_{t}+\frac{1}{2}f\partial_{r}=\frac{1}{2}f^{1/2}(u^{a}\partial_{a}+n^{a}\partial_{a}).
\end{equation}
For this choice, it is straightforward to work out the contraction
\begin{equation}
T_{ab}l^{a}l^{b}=0
\end{equation}
as well as the surface gravity and the 2-dimensional extrinsic curvature
\begin{equation}\label{RNnul-g2k}
g_{\mathcal{H}}=\frac{Mr-Q^2}{r^3},~~~~k_{AB}=\frac{f}{2r}\gamma_{AB}.
\end{equation}

Taking $\alpha=f^{1/2}$, when the stretched horizon becomes coincident with the true horizon $\alpha\rightarrow0$, from equations \eqref{RN3k}, \eqref{RNdu}, \eqref{RNnul-g2k} one can confirm relations \eqref{klimg}, \eqref{klimk}, \eqref{dulimk}, \eqref{klimdu}. With the expression of $\alpha$ unchanged, again in this case $\tilde{p}_{\mathcal{H}}=p_{\mathcal{H}}=(Mr-Q^2)/(8\pi Gr^3)$. In the EF coordinates and the KS coordinates respectively, the generator \eqref{lRN} can be reexpressed as
\begin{eqnarray}
\nonumber l^{a}\partial_{a}&=&\partial_{v}+\frac{1}{2}f\partial_{r}\\
&=&g_{+}x\partial_{x}.
\end{eqnarray}
Compared with \eqref{RN-EFnu} or \eqref{RN-KSnu}, it is evident that both $\alpha u^{a}$ and $\alpha n^{a}$ tend to $l^{a}$ in the near event horizon limit.

For the Schwarzschild black hole, in reference \cite{Price:1986yy} the event horizon's generator is chosen as $l^{a}\partial_{a}=\partial_{\bar{t}}$ with
\begin{equation}\label{lSch}
\bar{t}=t+\frac{1}{2g_{\mathcal{H}}}\ln(2g_{\mathcal{H}}\alpha^2)+\mathcal{O}(\alpha^2)
\end{equation}
up to a time translation. Our choice of the null generator in this subsection can be reformulated as $l^{a}\partial_{a}=\partial_{\bar{t}}$, where
\begin{eqnarray}
\nonumber \bar{t}&=&\frac{1}{g_{+}}\ln|x|\\
&=&t+r+\frac{1}{2g_{+}}\ln(r-r_{+})+\frac{1}{2g_{-}}\ln(r-r_{-}),
\end{eqnarray}
matching perfectly with the form \eqref{lSch} when $Q=0$.

\section{Membrane paradigm for dS universe}\label{sect-dS}
Ever since the establishment of membrane paradigm, attention has been focused on the horizon of black holes. However, as we will show in this and the next sections, the paradigm has a broader arena. It can be built for the cosmological horizon. Our purpose is to provide new examples of the application and the generality of the membrane paradigm, not development of a new paradigm. In this section, we will concentrate on the static case --- the dS universe. In the next section, we will deal with the FLRW universe.

For the $(3+1)$-dimensional dS spacetime with a positive cosmological constant $\Lambda$, the metric has a static form analogous to \eqref{RN},
\begin{equation}\label{dS}
ds^2=-f(r)d\tau^2+\frac{1}{f(r)}dr^2+r^2d\Omega^2
\end{equation}
with $f(r)=1-r^2/L^2$ and $L^2=3/\Lambda$. Here we denote the time coordinate with $\tau$, but save the notation $t$ for later use. The bulk stress tensor is
\begin{equation}
8\pi GT_{ab}dx^{a}dx^{b}=-\Lambda g_{ab}dx^{a}dx^{b}.
\end{equation}
The cosmological horizon of the above spacetime is an event horizon. Inside the horizon we have $f>0$. Our aim in this section is to construct a membrane paradigm for this horizon. Suppose slightly inside the cosmological horizon there is a stretched horizon with an inward-pointing spacelike unit normal $n^{a}\partial_{a}=-f^{1/2}\partial_{r}$ and a future-directed timelike generator $u^{a}\partial_{a}=f^{-1/2}\partial_{\tau}$, which imply
\begin{equation}
T_{ab}n^{a}u^{b}=0.
\end{equation}
The definition \eqref{3k} leads to the nonvanishing components of 3-dimensional extrinsic curvature
\begin{equation}
K_{\tau\tau}=-\frac{rf^{1/2}}{L^2},~~~~K_{\vartheta\vartheta}=-rf^{1/2},~~~~K_{\varphi\varphi}=-rf^{1/2}\sin^2\vartheta,
\end{equation}
from which we find
\begin{equation}\label{dS3k}
\alpha u^{a}u^{b}K_{ab}=-\frac{\alpha r}{L^2f^{1/2}},~~~~\alpha\gamma_{A}^{a}\gamma_{B}^{b}K_{ab}=-\frac{\alpha f^{1/2}}{r}\gamma_{AB},~~~~\gamma_{A}^{a}u^{b}K_{ab}=0.
\end{equation}
Straightforward calculation gives
\begin{equation}\label{dSdu}
\gamma_{A}^{a}\gamma_{B}^{b}\nabla_{b}(\alpha u_{a})=0,~~~~\gamma_{A}^{a}n^{b}\nabla_{b}u_{a}=0.
\end{equation}

Introducing the time coordinate $dt=d\tau-rL^{-1}f^{-1}dr$, we can rewrite the dS metric as
\begin{equation}\label{dS-EF}
ds^2=-dt^2+\left(dr-\frac{r}{L}dt\right)^2+r^2d\Omega^2,
\end{equation}
In this section, our calculation will be performed mainly in this coordinate system. In accord with the metric \eqref{dS-EF}, the unit normal and the timelike generator are rewritten as
\begin{eqnarray}\label{dS-EFnu}
\nonumber n^{a}\partial_{a}&=&rL^{-1}f^{-1/2}\partial_{t}-f^{1/2}\partial_{r},\\
u^{a}\partial_{a}&=&f^{-1/2}\partial_{t}.
\end{eqnarray}

For the dS spacetime, we can also define the KS-like coordinates $x$ and $y$ via
\begin{eqnarray}
\nonumber xy&=&-\frac{L-r}{L+r},\\
\frac{x}{y}&=&-e^{2\tau/L}.
\end{eqnarray}
In terms of them, the metric takes the form
\begin{equation}\label{dS-KS}
ds^2=-\frac{4L^2}{(1-xy)^2}dxdy+r^2d\Omega^2,
\end{equation}
while the unit normal and the timelike generator are transformed to
\begin{eqnarray}\label{dS-KSnu}
\nonumber n^{a}\partial_{a}&=&L^{-1}f^{-1/2}(x\partial_{x}+y\partial_{y}),\\
u^{a}\partial_{a}&=&L^{-1}f^{-1/2}(x\partial_{x}-y\partial_{y}).
\end{eqnarray}
The future event horizon is the $y=0$ hypersurface.

Facilitated with these results, now we are very close to our mission of building the dS membrane paradigm. The next step is specifying the generator of cosmological horizon. Like the RN black hole, the null condition and the Killing equation can be met together locally on the event horizon, but away from the horizon we should break either of them. Parallel to the previous section, we will investigate the dS horizon with a globally Killing generator in section \ref{subsect-dSKil} and a globally null generator in section \ref{subsect-dSnul}.

\subsection{Killing vector as horizon's generator}\label{subsect-dSKil}
In the static form, the dS metric has a Killing vector
\begin{equation}
l^{a}\partial_{a}=\partial_{\tau}=f^{1/2}u^{a}\partial_{a}
\end{equation}
which is null on the event horizon. Taking it as the global extension of the horizon's generator, we can see $\alpha=f^{1/2}$ and
\begin{equation}
T_{ab}l^{a}l^{b}=\frac{\Lambda}{8\pi G}f.
\end{equation}
The 2-dimensional extrinsic curvature is calculated according to definition \eqref{2k},
\begin{equation}\label{dSKil-2k}
k_{AB}=0.
\end{equation}
The surface gravity can be derived with formula \eqref{surfg} in metric \eqref{dS-EF},
\begin{equation}\label{dSKil-g}
g_{\mathcal{H}}=\frac{r^2}{L^3}.
\end{equation}

In the near event horizon limit $\alpha\rightarrow0$, relations \eqref{klimg}, \eqref{klimk}, \eqref{dulimk}, \eqref{klimdu} can be confirmed by using equations \eqref{dS3k}, \eqref{dSdu}, \eqref{dSKil-2k}, \eqref{dSKil-g}. Because $f$ is independent of $\tau$, in this case we find $u^{b}\nabla_{b}\alpha=0$ and $\tilde{p}_{\mathcal{H}}=p_{\mathcal{H}}=r^2/(8\pi GL^3)$ in the near horizon limit.

\subsection{Null vector as horizon's generator}\label{subsect-dSnul}
The other choice is extending the horizon's generator with the ingoing null vector
\begin{equation}\label{ldS}
l^{a}\partial_{a}=\partial_{t}-\left(1-\frac{r}{L}\right)\partial_{r}=f^{1/2}\left(1+\frac{r}{L}\right)^{-1}(u^{a}\partial_{a}+n^{a}\partial_{a}).
\end{equation}
For this choice, we can get the contraction
\begin{equation}
T_{ab}l^{a}l^{b}=0
\end{equation}
as well as the surface gravity and the 2-dimensional extrinsic curvature
\begin{equation}\label{dSnul-g2k}
g_{\mathcal{H}}=\frac{1}{L},~~~~k_{AB}=-\frac{1}{r}\left(1-\frac{r}{L}\right)\gamma_{AB}.
\end{equation}

In coordinate systems \eqref{dS} and \eqref{dS-KS} respectively, the generator \eqref{ldS} can be reexpressed as
\begin{eqnarray}
\nonumber l^{a}\partial_{a}&=&\left(1-\frac{r}{L}\right)(f^{-1}\partial_{\tau}-\partial_{r})\\
&=&\frac{x}{L}(1-xy)\partial_{x}.
\end{eqnarray}
Comparing it with \eqref{dS-EFnu} or \eqref{dS-KSnu}, we find $\alpha u^{a}\rightarrow l^{a}$ and $\alpha n^{a}\rightarrow l^{a}$ in the near event horizon limit if $\alpha=f^{1/2}$, and relations \eqref{klimg}, \eqref{klimk}, \eqref{dulimk}, \eqref{klimdu} are satisfactorily obtained from equations \eqref{dS3k}, \eqref{dSdu}, \eqref{dSnul-g2k}. Like the previous subsection, $u^{b}\nabla_{b}\alpha=0$ and thus $\tilde{p}_{\mathcal{H}}=p_{\mathcal{H}}=1/(8\pi GL)$ in the near horizon limit.

%\begin{eqnarray}
%\nonumber&&-\frac{1}{\alpha^2}p_{\mathcal{H}}\theta_{\mathcal{H}}-\rho_{\mathcal{H}}u^{b}\nabla_{b}\left(\frac{1}{\alpha}\right)\\
%\nonumber&=&-\frac{1}{\alpha^2}\theta_{\mathcal{H}}\left(p_{\mathcal{H}}+\frac{1}{8\pi G}\frac{1}{\alpha}l^{b}\nabla_{b}\alpha\right)\\
%&=&-\frac{1}{8\pi G}\frac{1}{\alpha^2}\theta_{\mathcal{H}}g_{\mathcal{H}}+T_{ab}\left(\frac{1}{\alpha^2}l^{a}l^{b}-n^{a}u^{b}\right).
%\end{eqnarray}

\section{Oblique membrane paradigm in FLRW universe}\label{sect-FLRW}
The familiar FLRW metric
\begin{equation}
ds^2=-dt^2+a^2\left(\frac{1}{1-k\tilde{r}^2}d\tilde{r}^2+\tilde{r}^2d\Omega^2\right)
\end{equation}
in $(3+1)$-dimensional spacetime can be transformed through $r=a\tilde{r}$ as\footnote{Be cautious that we have exchanged the notations $r$ and $\tilde{r}$ in comparison with reference \cite{Cai:2006rs}.}
\begin{equation}
ds^2=-dt^2+\frac{1}{1-ka^{-2}r^2}(dr-Hrdt)^2+r^2d\Omega^2.
\end{equation}
Written in the double-null coordinates \cite{Hayward:1993ph,Cai:2006rs}, it is
\begin{equation}
ds^2=-2d\xi^{+}d\xi^{-}+r^2d\Omega^2,
\end{equation}
where
\begin{equation}
d\xi^{\pm}=\frac{1}{\sqrt{2}}\left[\left(1\pm\frac{Hr}{\sqrt{1-ka^{-2}r^2}}\right)dt\mp\frac{1}{\sqrt{1-ka^{-2}r^2}}dr\right].
\end{equation}
The one-forms $-d\xi^{\mp}$ are dual to vectors
\begin{equation}
\partial_{\pm}\equiv\frac{\partial}{\partial\xi^{\pm}}=\frac{1}{\sqrt{2}}\partial_t+\frac{1}{\sqrt{2}}(Hr\mp\sqrt{1-ka^{-2}r^2})\partial_r.
\end{equation}
What we need in this section is the ingoing null vector $l^{a}\partial_{a}=\sqrt{2}\partial_{+}$ which can be put into the form
\begin{equation}\label{lFLRW}
l^{a}\partial_{a}=\partial_{t}-\left(\sqrt{1-ka^{-2}r^2}-Hr\right)\partial_{r}.
\end{equation}

For the FLRW universe, the trapping horizon $\sqrt{1-ka^{-2}r^2}-Hr=0$ coincides with the apparent horizon
\begin{equation}\label{traphor}
r^2=\frac{1}{H^2+ka^{-2}},
\end{equation}
whose normal vector can be written as
\begin{equation}\label{vFLRW}
v^{a}\partial_{a}=\left(1+\frac{\dot{H}}{H^2}\right)\partial_{t}+\frac{r}{H}\left(\dot{H}-ka^{-2}\right)\partial_{r}
\end{equation}
up to a normalization factor. On the trapping horizon, this vector can be normalized to
\begin{eqnarray}\label{vnorm}
\nonumber v_{a}v^{a}&=&-\frac{1}{H^4}\left(\dot{H}-ka^{-2}\right)\left(\dot{H}+2H^2+ka^{-2}\right)\\
&=&\frac{16\pi^2G^2}{3H^4}(\rho+p)(\rho-3p).
\end{eqnarray}
The cosmological equations \eqref{Friedmann1} and \eqref{Friedmann2} have been used in the second line. Therefore, the trapping horizon may be timelike, null or spacelike, depending on the equation of state. For all of the three cases, the membrane paradigm can be constructed by introducing a cosmological stretched horizon slightly inside the trapping horizon, which possesses an inward-pointing spacelike unit normal and a future-directed timelike generator\footnote{They can be obtained with the help of FLRW metric in the quasistatic coordinates \cite{YuWang} $ds^2=-(1-ka^{-2}r^2-H^2r^2)f^2d\tau^2+(1-ka^{-2}r^2-H^2r^2)^{-1}dr^2+r^2d\Omega^2$, where $n^{a}\partial_{a}=-\left(1-ka^{-2}r^2-H^2r^2\right)^{1/2}\partial_{r}$, $u^{a}\partial_{a}=\left(1-ka^{-2}r^2-H^2r^2\right)^{-1/2}f^{-1}\partial_{\tau}$.}
\begin{eqnarray}\label{FLRWnu}
\nonumber n^{a}\partial_{a}&=&Hr\left(1-ka^{-2}r^2-H^2r^2\right)^{-1/2}\partial_{t}-\left(1-ka^{-2}r^2-H^2r^2\right)^{1/2}\partial_{r},\\
u^{a}\partial_{a}&=&\left(1-ka^{-2}r^2\right)^{1/2}\left(1-ka^{-2}r^2-H^2r^2\right)^{-1/2}\partial_{t}
\end{eqnarray}
subject to orthonormal conditions \eqref{orth} and \eqref{norm}. They are related to the null vector \eqref{lFLRW} by
\begin{equation}
l^{a}\partial_{a}=\frac{\left(1-ka^{-2}r^2-H^2r^2\right)^{1/2}}{\sqrt{1-ka^{-2}r^2}+Hr}(u^{a}\partial_{a}+n^{a}\partial_{a}).
\end{equation}
In agreement with ansatz \eqref{ansatz}, their dual form can be expressed as $n_{a}=-N\nabla_{a}r$, $u_{a}=-Uh_{a}^{b}\nabla_{b}t$, where
\begin{eqnarray}
\nonumber N&=&\left(1-ka^{-2}r^2-H^2r^2\right)^{-1/2},\\
U&=&\left(1-ka^{-2}r^2\right)^{-1/2}\left(1-ka^{-2}r^2-H^2r^2\right)^{1/2}.
\end{eqnarray}
%$u_{a}=-U\nabla_{a}\tau$ with
%\begin{eqnarray}
%\nonumber&&Hr\partial_{t}\ln U-\left(1-ka^{-2}r^2-H^2r^2\right)\partial_{r}\ln U\\
%\nonumber&=&Hr\partial_{t}\ln\left(1-ka^{-2}r^2-H^2r^2\right)^{-1/2}\\
%\nonumber&&-\left(1-ka^{-2}r^2-H^2r^2\right)\partial_{r}\ln\left(1-ka^{-2}r^2-H^2r^2\right)^{-1/2}+\dot{H}r+ka^{-2}r+2H^2r\\
%\nonumber&=&Hr\partial_{t}\ln\left[Har\left(1-ka^{-2}r^2-H^2r^2\right)^{-1/2}\right]\\
%&&-\left(1-ka^{-2}r^2-H^2r^2\right)\partial_{r}\ln\left[Har\left(1-ka^{-2}r^2-H^2r^2\right)^{-1/2}\right]+\frac{1}{r}.
%\end{eqnarray}
Repeating calculations in the previous sections with the definition \eqref{3k}, we find the following components of the 3-dimensional extrinsic curvature are not vanishing
\begin{eqnarray}
\nonumber K_{tt}&=&-\frac{r\left[H^2\left(1-2ka^{-2}r^2-H^2r^2\right)+\dot{H}\left(1-ka^{-2}r^2\right)\right]}{\left(1-ka^{-2}r^2\right)\left(1-ka^{-2}r^2-H^2r^2\right)^{1/2}},\\
\nonumber K_{tr}&=&K_{rt}=-\frac{Hr^2\left[H^2\left(1-2ka^{-2}r^2-H^2r^2\right)+\dot{H}\left(1-ka^{-2}r^2\right)\right]}{\left(1-ka^{-2}r^2\right)\left(1-ka^{-2}r^2-H^2r^2\right)^{3/2}},\\
\nonumber K_{rr}&=&-\frac{H^2r^3\left[H^2\left(1-2ka^{-2}r^2-H^2r^2\right)+\dot{H}\left(1-ka^{-2}r^2\right)\right]}{\left(1-ka^{-2}r^2\right)\left(1-ka^{-2}r^2-H^2r^2\right)^{5/2}},\\
\nonumber K_{\vartheta\vartheta}&=&-r\left(1-ka^{-2}r^2-H^2r^2\right)^{1/2},\\
K_{\varphi\varphi}&=&-r\sin^2\vartheta\left(1-ka^{-2}r^2-H^2r^2\right)^{1/2}.
\end{eqnarray}
Projected to the temporal or transverse directions, they give rise to
\begin{eqnarray}\label{FLRW3k}
\nonumber\alpha u^{a}u^{b}K_{ab}&=&-\frac{\alpha r}{\left(1-ka^{-2}r^2-H^2r^2\right)^{3/2}}\left[H^2\left(1-2ka^{-2}r^2-H^2r^2\right)+\dot{H}\left(1-ka^{-2}r^2\right)\right],\\
\nonumber\alpha\gamma_{A}^{a}\gamma_{B}^{b}K_{ab}&=&-\frac{\alpha}{r}\left(1-ka^{-2}r^2-H^2r^2\right)^{1/2}\gamma_{AB},\\
\gamma_{A}^{a}u^{b}K_{ab}&=&0.
\end{eqnarray}
From the expressions \eqref{FLRWnu}, we can also get
\begin{equation}\label{FLRWdu}
\gamma_{A}^{a}\gamma_{B}^{b}\nabla_{b}(\alpha
u_{a})=0,~~~~\gamma_{A}^{a}n^{b}\nabla_{b}u_{a}=0.
\end{equation}

On the other hand, by definition \eqref{2k}, the 2-dimensional extrinsic curvature turns out to be
\begin{equation}\label{FLRW2k}
k_{AB}=-\frac{1}{r}\left(\sqrt{1-ka^{-2}r^2}-Hr\right)\gamma_{AB}.
\end{equation}
To recover the limit \eqref{klimk}, we require the renormalization parameter to be
\begin{equation}
\alpha=\frac{\left(1-ka^{-2}r^2-H^2r^2\right)^{1/2}}{\left(1-ka^{-2}r^2\right)^{1/2}}.
\end{equation}
It follows directly that $\alpha u^{a}\rightarrow l^{a}$, $\alpha n^{a}\rightarrow l^{a}$ in the near horizon limit $\alpha\rightarrow0$. However, unlike previous sections, the Lie derivative of $\alpha$ along the timelike generator is nonzero,
\begin{equation}\label{Lieud}
\mathcal{L}_{u}\alpha=-\frac{\dot{H}Hr^2}{1-ka^{-2}r^2-H^2r^2}+\frac{kr^2}{a^2}\left(\frac{H}{1-ka^{-2}r^2-H^2r^2}-\frac{H}{1-ka^{-2}r^2}\right).
\end{equation}
For this sake, the limit \eqref{klimg} is expected to be replaced by \eqref{klimgmod}. Hence in this section we will examine the limit \eqref{klimgmod} instead.

Note the null vector $l^{a}$ is neither the generator nor the normal of trapping horizon.\footnote{We are grateful to the referee for bringing this point into our attention.} This can be inferred from the product
\begin{eqnarray}\label{vl}
\nonumber v_{a}l^{a}&=&-\frac{1}{H^2}\left(\dot{H}-ka^{-2}\right)\\
&=&\frac{4\pi G}{H^2}(\rho+p)
\end{eqnarray}
which is nonvanishing unless $p=-\rho$. In this paper, the membrane paradigm is built with the null vector $l^{a}$ instead of the normal $v^{a}$ of the trapping horizon. Here is why. First, the Raychaudhuri equation \eqref{Ray1} and the H\'a\'{\j}i\v{c}ek equation \eqref{NS1} are derived for null geodesic congruences. Second, as we have mentioned before, $v^{a}$ may be spacelike, null or timelike, but $l^{a}$ can be utilized to give a unified form of membrane paradigm for all cases. Third, we find $v^{a}$ is intractable to build the membrane paradigm because of the nonvanishness of equation \eqref{Lieud}, while our choice of $l^{a}$ is validated by the success of this section and the next section.

In sections \ref{sect-RN} and \ref{sect-dS}, we have analyzed the membrane paradigm for the RN black hole horizon and the dS horizon with various forms of generator $l^{a}\partial_{a}$. It is confirmable that all of them obey the Killing equation
\begin{equation}\label{Kill}
\mathcal{L}_{l}g_{ab}=0
\end{equation}
on the event horizon. However, the apparent horizon of the FLRW universe is not a Killing horizon except for the dS subcase. This can be seen by substituting its normal vector \eqref{vFLRW} into the left hand side of \eqref{Kill},
\begin{equation}
\mathcal{L}_{v}g_{\vartheta\vartheta}=\frac{1}{\sin^2\vartheta}\mathcal{L}_{v}g_{\varphi\varphi}=\frac{2r^2}{H}\left(\dot{H}-ka^{-2}\right),
\end{equation}
which obviously violate the Killing equation. If instead we substitute the null vector \eqref{lFLRW} into the left hand side of \eqref{Kill}, we would find
\begin{equation}
\mathcal{L}_{l}g_{tr}=H\left(\frac{1}{\sqrt{1-ka^{-2}r^2}}-\frac{2Hr}{1-ka^{-2}r^2}\right),~~~~\mathcal{L}_{l}g_{rr}=\frac{2H}{1-ka^{-2}r^2},
\end{equation}
which violate the Killing equation again at the apparent horizon. As we have mentioned in section \ref{subsect-geom}, there are ambiguities in defining the surface gravity for non-Killing horizons. For the trapping horizon in the FLRW universe, if we implement the definition \eqref{surfg}, then its surface gravity is
\begin{equation}\label{FLRWg}
g_{\mathcal{H}}=H.
\end{equation}
Interestingly, this is in agreement with the temperature at the horizon in a pure dS spacetime \cite{Huang:2002qe}, though we are dealing now with more general spacetimes. Also note that the temperature $T=g_{\mathcal{H}}/(2\pi)$ is positive as long as the universe is expanding $H>0$.

Recalling the stress tensor
\begin{equation}
T_{ab}dx^{a}dx^{b}=\rho dt^2+\frac{p}{1-ka^{-2}r^2}(dr-Hrdt)^2+pr^2d\Omega^2,
\end{equation}
where $\rho$ and $p$ are density and pressure respectively of the bulk fluid, we obtain $T_{ab}l^{a}l^{b}=\rho+p$ and
\begin{equation}
T_{ab}n^{a}u^{b}=\frac{Hr\left(1-ka^{-2}r^2\right)^{1/2}(\rho+p)}{1-ka^{-2}r^2-H^2r^2}.
\end{equation}
This is consistent with the relation $T_{ab}l^{a}l^{b}=\alpha^2T_{ab}n^{a}u^{b}$ in the near trapping horizon limit \eqref{traphor}.

Making use of equations \eqref{FLRW3k}, \eqref{FLRWdu}, \eqref{FLRW2k}, \eqref{FLRWg} and taking the limit \eqref{traphor}, we have succeeded in reproducing \eqref{klimk}, \eqref{dulimk}, \eqref{klimdu}, but failed in getting \eqref{klimgmod} straightforwardly. After trials and errors, we finally discovered the following relation
\begin{equation}\label{klimgmmod}
\frac{1}{Hr}\left(1-ka^{-2}r^2\right)^{1/2}\alpha u^{a}u^{b}K_{ab}-\frac{1}{H^2r^2}\left(1-ka^{-2}r^2\right)\mathcal{L}_{u}\alpha=-g_{\mathcal{H}}.
\end{equation}
It recovers \eqref{Raylimcond2} or \eqref{klimgmod} when the stretched horizon approaches the trapping horizon in the limit \eqref{traphor}. This is reminiscent of results in previous sections. In section \ref{sect-RN}, it can be checked that the values of $\rho_{\mathcal{H}}$ and $\theta_{\mathcal{H}}$ meet condition \eqref{klimcond1} only in the near horizon limit. In section \ref{sect-dS}, the exact values of $p_{\mathcal{H}}$ and $\tilde{p}_{\mathcal{H}}$ become equal only at the apparent horizon.

Our $2+1+1$ split of the FLRW spacetime dictates a metric
\begin{equation}
\gamma_{AB}dx^{A}dx^{B}=r^2d\Omega^2,
\end{equation}
for the 2-dimensional spacelike section of the apparent horizon. Then it is easy to check $p_{\mathcal{H}||A}=\tilde{p}_{\mathcal{H}||A}=0$. This finishes our construction of the membrane paradigm for the apparent horizon in the FLRW universe. In this paradigm $\mathcal{L}_{u}\alpha\neq0$, so we dubbed it an ``oblique'' membrane paradigm. At the same time, the null vector $l^{a}$ is not always normal to the apprarent horizon, so this name is also suitable geometrically. By this name we do not mean a new paradigm. It is very clear from references \cite{Price:1986yy,Thorne86,Parikh:1997ma} that the so called pressure of the fluid depends upon the choice of the generators. It agrees with the surface gravity only in the case that the horizon is stationary. Here we have rediscovered this with an explicit example (FLRW). In section \ref{sect-stand}, we will discuss the standard membrane paradigm that is not oblique.

\section{Fluid equations to Friedmann equation}\label{sect-Friedmann}
It ought not to be surprising that the Friedmann equation emerges naturally from dynamics of membrane paradigm. There are two approaches to this end.

First, from the viewpoint of the stretched horizon, the membrane can be treated as a 2-dimensional fluid characterized by the quantities
\begin{eqnarray}
\nonumber&&\rho_{\mathcal{S}}=-\frac{\theta_{\mathcal{S}}}{8\pi G}=\frac{1}{4\pi Gr}\left(1-ka^{-2}r^2-H^2r^2\right)^{1/2},\\
\nonumber&&p_{\mathcal{S}}=\frac{r}{8\pi G\left(1-ka^{-2}r^2-H^2r^2\right)^{3/2}}\left[H^2\left(1-2ka^{-2}r^2-H^2r^2\right)+\dot{H}\left(1-ka^{-2}r^2\right)\right],\\
&&\sigma_{\mathcal{S}}^{AB}=0,~~~~\pi_{\mathcal{S}}^{A}=0,~~~~\eta_{\mathcal{S}}=\frac{1}{16\pi G},~~~~\zeta_{\mathcal{S}}=-\frac{1}{16\pi G}.
\end{eqnarray}
These quantities automatically fulfill equation \eqref{TnA}, that is
\begin{eqnarray}
\nonumber-T_{a}^{c}n_{c}\gamma_{A}^{a}&=&[\gamma_{AA'}\gamma_{B}^{b}(p_{\mathcal{S}}\gamma^{A'B}-2\eta_{\mathcal{S}}\sigma_{\mathcal{S}}^{A'B}-\zeta_{\mathcal{S}}\theta_{\mathcal{S}}\gamma^{A'B})]_{||b}\\
&&+\pi_{\mathcal{S}A}\gamma_{b}^{c}\nabla_{c}u^{b}+\gamma_{A}^{e}u^{c}\nabla_{c}\pi_{\mathcal{S}e}+\pi_{\mathcal{S}}^{d}\gamma_{A}^{e}\nabla_{d}u_{e}.
\end{eqnarray}
They should also satisfy equation \eqref{Tnu}, namely
\begin{equation}
-T_{b}^{a}n_{a}u^{b}=-(p_{\mathcal{S}}\gamma^{AB}-2\eta_{\mathcal{S}}\sigma_{\mathcal{S}}^{AB}-\zeta_{\mathcal{S}}\theta_{\mathcal{S}}\gamma^{AB})\gamma_{A}^{a}\gamma_{B}^{b}\nabla_{b}u_{a}-\nabla_{b}\pi_{\mathcal{S}}^{b}-u^{b}\nabla_{b}\rho_{\mathcal{S}}-\rho_{\mathcal{S}}\gamma_{b}^{a}\nabla_{a}u^{b},
\end{equation}
which turns out to be one of the cosmological equations in the FLRW universe
\begin{equation}\label{Friedmann1}
\dot{H}-\frac{k}{a^2}=-4\pi G(\rho+p).
\end{equation}
In this paper, the dot overhead implies the derivative with respect to $t$, and the superscript prime denotes the derivative with respect to $r$ if not in the indices. Given the continuity equation of the bulk fluid $\dot{\rho}+3H(\rho+p)=0$, the above equation can be integrated directly, yielding the Friedmann equation
\begin{equation}\label{Friedmann2}
H^2+\frac{k}{a^2}=\frac{8\pi G}{3}\rho.
\end{equation}

Second, in the near horizon limit, we can also consider the 2-dimensional fluid on the apparent horizon, endowed with the fluid quantities
\begin{eqnarray}
\nonumber&&\rho_{\mathcal{H}}=-\frac{\theta_{\mathcal{H}}}{8\pi G}=\frac{1}{4\pi Gr}\left(\sqrt{1-ka^{-2}r^2}-Hr\right),~~~~\tilde{p}_{\mathcal{H}}=\frac{H}{8\pi G},\\
&&\sigma_{\mathcal{H}}^{AB}=0,~~~~\pi_{\mathcal{H}}^{A}=0,~~~~\eta_{\mathcal{H}}=\frac{1}{16\pi G},~~~~\zeta_{\mathcal{H}}=-\frac{1}{16\pi G}.
\end{eqnarray}
Substituted into the continuity equation \eqref{Ray2} of the membrane fluid, it recovers equation \eqref{Friedmann1} and hence can be integrated to give the Friedmann equation \eqref{Friedmann2}. The Navier-Stokes equation \eqref{NS2} is automatically satisfied by the above fluid quantities on the apparent horizon.

Dividing the shear viscosity by the Bekenstein-Hawking entropy density $s=1/4G$, we find the ratio of the shear viscosity to the entropy density for the membrane fluid is
\begin{equation}
\frac{\eta_{\mathcal{S},\mathcal{H}}}{s}=\frac{1}{4\pi}
\end{equation}
on both the stretched and the apparent horizons.

\section{Towards standard membrane paradigm if there is}\label{sect-stand}
In the standard membrane paradigm for black holes, the fluid pressure $p_{\mathcal{S}}$ on the stretched horizon tends to the effective pressure $\tilde{p}_{\mathcal{H}}$ on the event horizon up to a renormalization parameter $\alpha$. For the trapping horizon in the FLRW universe, in section \ref{sect-FLRW} there is a membrane paradigm with $\mathcal{L}_{u}\alpha\neq0$, dubbed an oblique membrane paradigm, which is a specific example of the existing membrane paradigm. As we have demonstrated in section \ref{sect-rev}, the vanishness of $\mathcal{L}_{u}\alpha$ determines the the equality of $\alpha p_{\mathcal{S}}$ and $\tilde{p}_{\mathcal{H}}$ in the near true horizon limit. Therefore, the vanishness of $\mathcal{L}_{u}\alpha$ is a criterion for whether the membrane paradigm is standard or oblique.

One may wonder, in the FLRW universe, if there is a standard membrane paradigm that is not oblique, which is also a specific example of the existing membrane paradigm. Perhaps this can be devised by calibrating the normal and the timelike directions of the stretched horizon, the null generator of the trapping horizon and the renormalization parameter. In fact, it is very clear from the original derivation of the black-hole membrane paradigm or the derivation a la Parikh-Wilczek that the so called pressure of the fluid depends upon the choice of the generators. We have spent a long time on this problem but failed to get an answer. Here are partial results that may be helpful for future investigations. For conciseness, we restrict to the spatially flat case $k=0$ and denote $f=1-H^2r^2$.

Keeping the orthonormal properties, we can make a Bogoliubov transformation for vectors \eqref{FLRWnu},
\begin{eqnarray}
\nonumber\tilde{n}^{a}\partial_{a}&=&\sqrt{1+A^2}n^{a}\partial_{a}+Au^{a}\partial_{a}\\
\nonumber&=&f^{-1/2}\left(A+Hr\sqrt{1+A^2}\right)\partial_{t}-f^{1/2}\sqrt{1+A^2}\partial_{r},\\
\nonumber\tilde{u}^{a}\partial_{a}&=&An^{a}\partial_{a}+\sqrt{1+A^2}u^{a}\partial_{a}\\
&=&f^{-1/2}\left(\sqrt{1+A^2}+HrA\right)\partial_{t}-f^{1/2}A\partial_{r}.
\end{eqnarray}
We are still free to rescale the null generator \eqref{lFLRW},
\begin{eqnarray}
\nonumber\tilde{l}^{a}\partial_{a}&=&Bl^{a}\partial_{a}\\
\nonumber&=&B\partial_{t}-B(1-Hr)\partial_{r}\\
%\nonumber&=&\frac{Bf^{1/2}}{1+Hr}(n^{a}\partial_{a}+u^{a}\partial_{a})\\
&=&\frac{Bf^{1/2}}{1+Hr}\frac{\tilde{n}^{a}\partial_{a}+\tilde{u}^{a}\partial_{a}}{A+\sqrt{1+A^2}}.
\end{eqnarray}
The renormalization parameter is chosen as
\begin{equation}
\alpha=\frac{CBf^{1/2}}{A+\sqrt{1+A^2}}
\end{equation}
so that $C\rightarrow1$, $\alpha u^{a}\rightarrow l^{a}$, $\alpha n^{a}\rightarrow l^{a}$ in the near trapping horizon limit $Hr\rightarrow1$. Calculation from $\tilde{n}^{a}$, $\tilde{u}^{a}$ and $\tilde{l}^{a}$ yields
\begin{eqnarray}
\nonumber\alpha\tilde{u}^{a}\tilde{u}^{b}K_{ab}&=&-\frac{\alpha}{\sqrt{1+A^2}f^{1/2}}\Bigl[\left(\sqrt{1+A^2}+HrA\right)\left(\dot{A}+f^{-1}\dot{H}r\sqrt{1+A^2}\right)\\
\nonumber&&+H^2r(1+A^2)-AA'f\Bigr]\\
\nonumber&=&-\frac{\alpha}{f^{1/2}}\left(\sqrt{1+A^2}+HrA\right)\partial_{t}\ln\left[\left(A+\sqrt{1+A^2}\right)\sqrt{\frac{1+Hr}{1-Hr}}\right]\\
\nonumber&&+\alpha\partial_{r}\left(f^{1/2}\sqrt{1+A^2}\right),\\
\nonumber g_{\mathcal{H}}&=&HB+\dot{B}-B'(1-Hr)\\
&=&\partial_{t}B-\partial_{r}[B(1-Hr)].
\end{eqnarray}
The search of a standard membrane paradigm becomes now the search of functions $A$ and $B$ meeting conditions
\begin{equation}
\lim_{Hr\rightarrow1}g_{\mathcal{H}}+\alpha u^{a}u^{b}K_{ab}=0,~~~~\lim_{Hr\rightarrow1}\alpha=0.
\end{equation}
Unfortunately, we can neither find suitable functions nor disprove strictly their existence.
%\begin{equation}
%B=\left(A+\sqrt{1+A^2}\right)\sqrt{\frac{1+Hr}{1-Hr}}.
%\end{equation}

\section{Conclusion}\label{sect-con}
In this paper, we reanalyzed the membrane paradigm of black holes with the action principle \cite{Parikh:1997ma}, and rediscovered the membrane can be oblique if the derivative of renormalization parameter is nonvanishing along the timelike generator.

The standard membrane paradigm was realized concretely for the RN black hole. Although the event horizon of RN black hole is a null Killing horizon, outside the horizon we have to violate either the null condition or the Killing equation. Therefore we extended the horizon's generator to the full spacetime with a globally Killing generator in section \ref{subsect-RNKil} and a globally null generator in section \ref{subsect-RNnul} respectively, and confirmed that the membrane paradigm goes well.

As a new example, we established a standard membrane paradigm near the event horizon of dS universe, with a globally Killing generator in section \ref{subsect-dSKil} and a globally null generator in section \ref{subsect-dSnul}. In this example, the stretched horizon is located inside the event horizon, and the spacelike normal vector of the stretched horizon is pointing inward.

The most tantalizing part is section \ref{sect-FLRW}, in which we applied the membrane paradigm to the apparent horizon of FLRW spacetime. Because the renormalization parameter is time-dependent, its derivative along the time direction is nonzero and contributes a correction to the effective pressure on the apparent horizon. In convenient words, the stretched horizon is oblique in this paradigm, which is another example of the existing membrane paradigm \cite{Price:1986yy,Thorne86,Parikh:1997ma}. We derived the cosmological equations from the fluid dynamics in the membrane paradigm.

We also put forward a method to the standard membrane paradigm that is not oblique for the FLRW universe, if there is such a paradigm. We have not get a conclusive answer, but leave it as an open problem to strong researchers for future investigation.

In the near future, we wish to connect the membrane paradigm to observational cosmology, such as cosmic inflation and accelerated expansion.

\begin{acknowledgments}
This work is supported by the Science and Technology Commission of Shanghai Municipality (Grant No. 11DZ2260700), and partly by the National Natural Science Foundation of China (Grant No. 11105053). Preliminary results of this paper have been presented in 2014 Fall Meeting of the Chinese Physical Society. The author would like to thank Qing Wang, Jun-Bao Wu and other participants for illuminating comments. The author is also grateful to referees for suggestions on improving the manuscript.
\end{acknowledgments}

\appendix

\section{Derivation of fluid equations}\label{app-der}
%\section{Deriving fluid equations on the membrane}\label{app-deriv}
First we note that the orthogonal relation \eqref{orth} yields
\begin{eqnarray}
\nonumber&&u_{b}\gamma^{ab}=n_{b}\gamma^{ab}=n_{b}h^{ab}=0,~~~~u_{b}g^{bc}=u_{b}h^{bc},\\
\nonumber&&\gamma_{ab}g^{bc}=\gamma_{ab}h^{bc}=\gamma_{ab}\gamma^{bc},~~~~h_{ab}g^{bc}=h_{ab}h^{bc},\\
&&h_{a}^{b}h_{b}^{c}=h_{a}^{c},~~~~\gamma_{a}^{b}\gamma_{b}^{c}=\gamma_{a}^{c}.
\end{eqnarray}

Second, from normalization conditions \eqref{norm}, it is easy to see
\begin{equation}
n^{a}\nabla_{b}n_{a}=u^{a}\nabla_{b}u_{a}=0.
\end{equation}

Third, without loss of generality, we parameterize the the spacelike normal vector with an affine or non-affine parameter $\lambda$ by $n_{a}=N\nabla_{a}\lambda$, and the timelike generator with $\tau$ by $u_{a}=-Uh_{a}^{b}\nabla_{b}\tau$. Such a parametrization guarantees the orthogonal condition \eqref{orth}. Then it is straightforward to prove
\begin{eqnarray}
\nonumber n^{a}\nabla_{a}n^{b}&=&(N\nabla^{a}\lambda)\nabla_{a}(N\nabla^{b}\lambda)\\
\nonumber&=&(N\nabla^{a}\lambda)(\nabla_{a}N)\nabla^{b}\lambda+(N\nabla^{a}\lambda)(N\nabla^{b}\nabla_{a}\lambda)\\
\nonumber&=&n^{a}n^{b}\nabla_{a}\ln N+\frac{1}{2}N^2\nabla^{b}[(\nabla^{a}\lambda)(\nabla_{a}\lambda)]\\
\nonumber&=&n^{a}n^{b}\nabla_{a}\ln N+\frac{1}{2}N^2\nabla^{b}N^{-2}\\
\nonumber&=&-h^{ab}\nabla_{a}\ln N\\
&=&-\gamma^{ab}\nabla_{a}\ln N+u^{b}u^{a}\nabla_{a}\ln N,
\end{eqnarray}
\begin{eqnarray}
\nonumber u^{a}\nabla_{a}u^{b}&=&(Uh^{ac}\nabla_{c}\tau)\nabla_{a}(Uh^{bd}\nabla_{d}\tau)\\
\nonumber&=&(Uh^{ac}\nabla_{c}\tau)(\nabla_{a}U)h^{bd}\nabla_{d}\tau+(Uh^{ac}\nabla_{c}\tau)U[\nabla_{a}(g^{bd}-n^{b}n^{d})]\nabla_{d}\tau\\
\nonumber&&+(Uh^{ac}\nabla_{c}\tau)Uh^{bd}\nabla_{d}\nabla_{a}\tau\\
%\nonumber&=&u^{a}u^{b}\nabla_{a}\ln U-(Uh^{ac}\nabla_{c}\tau)U(n^{d}\nabla_{a}n^{b}+n^{b}\nabla_{a}n^{d})\nabla_{d}\tau\\
%\nonumber&&+\frac{1}{2}U^2h^{bd}\nabla_{d}[h^{ac}(\nabla_{a}\tau)(\nabla_{c}\tau)]-\frac{1}{2}U^2h^{bd}[\nabla_{d}(g^{ac}-n^{a}n^{c})](\nabla_{a}\tau)(\nabla_{c}\tau)\\
\nonumber&=&u^{a}u^{b}\nabla_{a}\ln U+u^{a}U(\underline{n^{d}}\nabla_{a}n^{b}+n^{b}\underline{\underline{\nabla_{a}n^{d}}})(\underline{h_{d}^{c}}+\underline{\underline{n_{d}}}n^{c})\nabla_{c}\tau\\
\nonumber&&+\frac{1}{2}U^2h^{bd}\nabla_{d}[h^{ac}(\nabla_{a}\tau)(\nabla_{c}\tau)]\\
\nonumber&&+\frac{1}{2}U^2h^{bd}(n^{c}\underline{\nabla_{d}n^{a}}+\underline{\underline{n^{a}}}\nabla_{d}n^{c})(\underline{\underline{h_{a}^{e}}}+\underline{n_{a}}n^{e})(\nabla_{e}\tau)(\nabla_{c}\tau)\\
\nonumber&=&u^{a}u^{b}\nabla_{a}\ln U+u_{e}h^{ea}U(n^{c}\nabla_{a}n^{b}+n^{b}h_{d}^{c}\nabla_{a}n^{d})\nabla_{c}\tau\\
\nonumber&&+\frac{1}{2}U^2h^{bd}\nabla_{d}(-U^{-2})+\frac{1}{2}U^2h^{bd}(n^{c}h_{a}^{e}\nabla_{d}n^{a}+n^{e}\nabla_{d}n^{c})(\nabla_{e}\tau)(\nabla_{c}\tau)\\
\nonumber&=&(h^{ab}+u^{a}u^{b})\nabla_{a}\ln U+u_{e}U(n^{c}K^{eb}+n^{b}K^{ec})\nabla_{c}\tau\\
\nonumber&&+\frac{1}{2}U^2(n^{c}K^{be}+n^{e}K^{bc})(\nabla_{e}\tau)(\nabla_{c}\tau)\\
\nonumber&=&\gamma^{ab}\nabla_{a}\ln U+u_{e}Un^{c}K^{eb}\nabla_{c}\tau-u_{e}u_{c}n^{b}K^{ec}\\
\nonumber&&-\frac{1}{2}u_{e}Un^{c}K^{be}\nabla_{c}\tau-\frac{1}{2}u_{c}Un^{e}K^{bc}\nabla_{e}\tau\\
&=&\gamma^{ab}\nabla_{a}\ln U-n^{b}K_{ac}u^{a}u^{c}.
\end{eqnarray}
For the membrane paradigm of RN black holes and the paradigm of cosmological horizons, it can be confirmed that $\gamma^{ab}\nabla_{a}\ln N=\gamma^{ab}\nabla_{a}\ln U=0$. Therefore, throughout this paper, it is safe make the ansatz \eqref{ansatz}. In the literature, a simple relation $n^{a}\nabla_{a}n^{b}=0$ is usually assumed, see e.g. \cite{Parikh:1997ma,Chatterjee:2010gp}. This is true for black holes because $u^{a}\nabla_{a}\ln N=0$. For the FLRW universe, one should be cautious that the simple relation does not hold any more. Ansatz \eqref{ansatz} is more robust than the old ansatz $n^{a}\nabla_{a}n^{b}=0$ in our first version of manuscript in e-Print archive, although it does not modify the final result of \eqref{Tnu} because of the wonderful cancellation in the last step.

The decomposition of stress tensor \eqref{mem-stress} can be recast equivalently
\begin{eqnarray}
\nonumber&&t_{\mathcal{S}}^{ab}\gamma_{a}^{A}u_{b}=t_{\mathcal{S}}^{ab}\gamma_{b}^{A}u_{a}=-\pi_{\mathcal{H}}^{A},~~~~t_{\mathcal{S}}^{ab}u_{a}u_{b}=\frac{1}{\alpha}\rho_{\mathcal{H}},\\
&&t_{\mathcal{S}}^{ab}\gamma_{a}^{A}\gamma_{b}^{B}=\frac{1}{\alpha}(p_{\mathcal{H}}\gamma^{AB}-2\eta_{\mathcal{H}}\sigma_{\mathcal{H}}^{AB}-\zeta_{\mathcal{H}}\theta_{\mathcal{H}}\gamma^{AB}),
\end{eqnarray}
which imply
%\begin{equation}
%n_{b}t_{\mathcal{S}}^{ab}=0,~~~~t_{\mathcal{S}}^{ab}g_{bc}=t_{\mathcal{S}}^{ab}h_{bc}
%\end{equation}
\begin{eqnarray}
\nonumber u_{b}\sigma_{\mathcal{H}}^{ab}=n_{b}\sigma_{\mathcal{H}}^{ab}=0,&&\sigma_{\mathcal{H}}^{ab}g_{bc}=\sigma_{\mathcal{H}}^{ab}h_{bc}=\sigma_{\mathcal{H}}^{ab}\gamma_{bc},\\
\nonumber u_{b}\pi_{\mathcal{H}}^{b}=n_{b}\pi_{\mathcal{H}}^{b}=0,&&\pi_{\mathcal{H}}^{b}g_{bc}=\pi_{\mathcal{H}}^{b}h_{bc}=\pi_{\mathcal{H}}^{b}\gamma_{bc},\\
n_{b}t_{\mathcal{S}}^{ab}=0,&&t_{\mathcal{S}}^{ab}g_{bc}=t_{\mathcal{S}}^{ab}h_{bc}.
\end{eqnarray}

Starting with equations \eqref{mem-stress} and \eqref{Gauss-Codazzi}, we can prove
\begin{eqnarray}\label{Tnu}
\nonumber-T_{b}^{a}n_{a}u^{b}&=&u_{a}t_{\mathcal{S}|b}^{ab}\\
\nonumber&=&u_{a}h_{d}^{a}(g_{b}^{c}-n_{b}n^{c})\nabla_{c}t_{\mathcal{S}}^{db}\\
\nonumber&=&u_{d}\nabla_{b}t_{\mathcal{S}}^{db}-u_{d}n^{c}\nabla_{c}(\underline{n_{b}t_{\mathcal{S}}^{db}})+u_{d}t_{\mathcal{S}}^{db}n^{c}\nabla_{c}n_{b}\\
\nonumber&=&u_{a}\nabla_{b}t_{\mathcal{S}}^{ab}+u_{d}t_{\mathcal{S}}^{db}n^{c}\nabla_{c}n_{b}\\
\nonumber&=&u_{a}\nabla_{b}[t_{\mathcal{S}}^{cd}(\gamma_{c}^{a}-u_{c}u^{a})(\gamma_{d}^{b}-u_{d}u^{b})]+u_{d}t_{\mathcal{S}}^{db}n^{c}\nabla_{c}n_{b}\\
\nonumber&=&u_{a}\nabla_{b}(t_{\mathcal{S}}^{cd}\gamma_{c}^{a}\gamma_{d}^{b}-t_{\mathcal{S}}^{cd}\gamma_{c}^{a}u_{d}u^{b}-t_{\mathcal{S}}^{cd}\gamma_{d}^{b}u_{c}u^{a}+t_{\mathcal{S}}^{cd}u_{c}u^{a}u_{d}u^{b})+u_{d}t_{\mathcal{S}}^{db}n^{c}\nabla_{c}n_{b}\\
\nonumber&=&\nabla_{b}(\underline{u_{a}}t_{\mathcal{S}}^{cd}\underline{\gamma_{c}^{a}}\gamma_{d}^{b})-t_{\mathcal{S}}^{cd}\gamma_{c}^{a}\gamma_{d}^{b}\nabla_{b}u_{a}-\nabla_{b}(\underline{u_{a}}t_{\mathcal{S}}^{cd}\underline{\gamma_{c}^{a}}u_{d}u^{b})\\
\nonumber&&+t_{\mathcal{S}}^{cd}\underline{\gamma_{c}^{a}}u_{d}\underline{u^{b}\nabla_{b}u_{a}}-\nabla_{b}(u_{a}t_{\mathcal{S}}^{cd}\gamma_{d}^{b}u_{c}u^{a})+t_{\mathcal{S}}^{cd}\gamma_{d}^{b}u_{c}\underline{u^{a}\nabla_{b}u_{a}}\\
\nonumber&&+\nabla_{b}(u_{a}t_{\mathcal{S}}^{cd}u_{c}u^{a}u_{d}u^{b})-t_{\mathcal{S}}^{cd}u_{c}\underline{u^{a}}u_{d}u^{b}\underline{\nabla_{b}u_{a}}+u_{d}t_{\mathcal{S}}^{db}n^{c}\nabla_{c}n_{b}\\
\nonumber&=&-t_{\mathcal{S}}^{cd}\gamma_{c}^{a}\gamma_{d}^{b}\nabla_{b}u_{a}+\nabla_{b}(t_{\mathcal{S}}^{cd}\gamma_{d}^{b}u_{c})-\nabla_{b}(t_{\mathcal{S}}^{cd}u_{c}u_{d}u^{b})+u_{d}t_{\mathcal{S}}^{db}n^{c}\nabla_{c}n_{b}\\
\nonumber&=&-t_{\mathcal{S}}^{cd}\gamma_{c}^{a}\gamma_{d}^{b}\nabla_{b}u_{a}+\nabla_{b}(t_{\mathcal{S}}^{cd}\gamma_{d}^{b}u_{c})-u^{b}\nabla_{b}(t_{\mathcal{S}}^{cd}u_{c}u_{d})\\
\nonumber&&-t_{\mathcal{S}}^{cd}u_{c}u_{d}(\gamma_{b}^{a}-\underline{u_{b}}u^{a}+n_{b}n^{a})\underline{\nabla_{a}u^{b}}+u_{d}t_{\mathcal{S}}^{db}n^{c}\nabla_{c}n_{b}\\
\nonumber&=&-t_{\mathcal{S}}^{cd}\gamma_{c}^{a}\gamma_{d}^{b}\nabla_{b}u_{a}+\nabla_{b}(t_{\mathcal{S}}^{cd}\gamma_{d}^{b}u_{c})-u^{b}\nabla_{b}(t_{\mathcal{S}}^{cd}u_{c}u_{d})-t_{\mathcal{S}}^{cd}u_{c}u_{d}\gamma_{b}^{a}\nabla_{a}u^{b}\\
\nonumber&&+\underline{t_{\mathcal{S}}^{cd}u_{c}u_{d}u^{b}n^{a}\nabla_{a}n_{b}+u_{d}t_{\mathcal{S}}^{db}n^{c}\nabla_{c}n_{b}}\\
&=&-t_{\mathcal{S}}^{cd}\gamma_{c}^{a}\gamma_{d}^{b}\nabla_{b}u_{a}+\nabla_{b}(t_{\mathcal{S}}^{cd}\gamma_{d}^{b}u_{c})-u^{b}\nabla_{b}(t_{\mathcal{S}}^{cd}u_{c}u_{d})-t_{\mathcal{S}}^{cd}u_{c}u_{d}\gamma_{b}^{a}\nabla_{a}u^{b}.
\end{eqnarray}
When the stretched horizon approaches the true horizon $\alpha\rightarrow0$, both $\alpha u^{a}$ and $\alpha n^{a}$ approach $l^{a}$, and the above equation can be written as
\begin{eqnarray}
\nonumber-\frac{1}{\alpha^2}T_{b}^{a}l_{a}l^{b}&=&-\frac{1}{\alpha}(p_{\mathcal{H}}\gamma^{AB}-2\eta_{\mathcal{H}}\sigma_{\mathcal{H}}^{AB}-\zeta_{\mathcal{H}}\theta_{\mathcal{H}}\gamma^{AB})\gamma_{A}^{a}\gamma_{B}^{b}\nabla_{b}u_{a}-\nabla_{b}\pi_{\mathcal{H}}^{b}\\
\nonumber&&-u^{b}\nabla_{b}\left(\frac{\rho_{\mathcal{H}}}{\alpha}\right)-\frac{1}{\alpha}\rho_{\mathcal{H}}\gamma_{b}^{a}\nabla_{a}u^{b}\\
\nonumber&=&-\frac{1}{\alpha^2}(p_{\mathcal{H}}\gamma^{AB}-2\eta_{\mathcal{H}}\sigma_{\mathcal{H}}^{AB}-\zeta_{\mathcal{H}}\theta_{\mathcal{H}}\gamma^{AB})\gamma_{A}^{a}\gamma_{B}^{b}\nabla_{b}(\alpha u_{a})-\nabla_{b}\pi_{\mathcal{H}}^{b}\\
\nonumber&&-\frac{1}{\alpha}u^{b}\nabla_{b}\rho_{\mathcal{H}}-\rho_{\mathcal{H}}u^{b}\nabla_{b}\left(\frac{1}{\alpha}\right)-\frac{1}{\alpha^2}\rho_{\mathcal{H}}\gamma_{b}^{a}\nabla_{a}(\alpha u^{b})\\
\nonumber&=&-\frac{1}{\alpha^2}(p_{\mathcal{H}}\gamma^{AB}-2\eta_{\mathcal{H}}\sigma_{\mathcal{H}}^{AB}-\zeta_{\mathcal{H}}\theta_{\mathcal{H}}\gamma^{AB})\gamma_{A}^{a}\gamma_{B}^{b}\nabla_{b}l_{a}-\nabla_{b}\pi_{\mathcal{H}}^{b}\\
\nonumber&&-\frac{1}{\alpha^2}l^{b}\nabla_{b}\rho_{\mathcal{H}}+\frac{1}{\alpha^2}\rho_{\mathcal{H}}u^{b}\nabla_{b}\alpha-\frac{1}{\alpha^2}\rho_{\mathcal{H}}\gamma_{b}^{a}\nabla_{a}l^{b}\\
\nonumber&=&-\frac{1}{\alpha^2}(p_{\mathcal{H}}\gamma^{AB}-2\eta_{\mathcal{H}}\sigma_{\mathcal{H}}^{AB}-\zeta_{\mathcal{H}}\theta_{\mathcal{H}}\gamma^{AB})k_{AB}-\nabla_{b}\pi_{\mathcal{H}}^{b}\\
&&-\frac{1}{\alpha^2}l^{b}\nabla_{b}\rho_{\mathcal{H}}+\frac{1}{\alpha^2}\rho_{\mathcal{H}}u^{b}\nabla_{b}\alpha-\frac{1}{\alpha^2}\rho_{\mathcal{H}}\gamma_{b}^{a}k_{a}^{b}.
\end{eqnarray}

Similarly, we can start with equations \eqref{mem-stress} and \eqref{Gauss-Codazzi} to prove
\begin{eqnarray}\label{TnA}
\nonumber-T_{a}^{c}n_{c}\gamma_{A}^{a}&=&\gamma_{aA}t_{\mathcal{S}|b}^{ab}\\
\nonumber&=&(\gamma_{aA}t_{\mathcal{S}}^{ab})_{|b}-t_{\mathcal{S}}^{ab}\gamma_{aA|b}\\
\nonumber&=&h_{b}^{c}h_{A}^{e}\nabla_{c}(\gamma_{ae}t_{\mathcal{S}}^{ab})-t_{\mathcal{S}}^{ab}(\underline{h_{aA}}+u_{a}u_{A})\underline{_{|b}}\\
\nonumber&=&(\gamma_{b}^{c}\gamma_{A}^{e}-\gamma_{b}^{c}u_{A}u^{e}-u_{b}u^{c}h_{A}^{e})\nabla_{c}(\gamma_{ae}t_{\mathcal{S}}^{ab})-t_{\mathcal{S}}^{ab}h_{a}^{c}h_{A}^{e}h_{b}^{d}\nabla_{d}(u_{c}u_{e})\\
\nonumber&=&\gamma_{b}^{c}\gamma_{A}^{e}\nabla_{c}[\gamma_{ae}t_{\mathcal{S}}^{ad}(\gamma_{d}^{b}-u_{d}u^{b})]-\gamma_{b}^{c}u_{A}\nabla_{c}(\underline{u^{e}\gamma_{ae}}t_{\mathcal{S}}^{ab})+\gamma_{b}^{c}\gamma_{ae}t_{\mathcal{S}}^{ab}u_{A}\nabla_{c}u^{e}\\
\nonumber&&-u^{c}h_{A}^{e}\nabla_{c}(u_{b}\gamma_{ae}t_{\mathcal{S}}^{ab})+\underline{u^{c}}h_{A}^{e}\gamma_{ae}\underline{t_{\mathcal{S}}^{ab}\nabla_{c}u_{b}}-t_{\mathcal{S}}^{cd}h_{A}^{e}\nabla_{d}(u_{c}u_{e})\\
\nonumber&=&(\gamma_{aA}t_{\mathcal{S}}^{ad}\gamma_{d}^{b})_{||b}-\gamma_{b}^{c}\gamma_{A}^{e}\nabla_{c}(\gamma_{ae}t_{\mathcal{S}}^{ad}u_{d}u^{b})\\
\nonumber&&+(h_{b}^{c}+u_{b}\underline{u^{c}})(h_{ae}+u_{a}\underline{u_{e}})t_{\mathcal{S}}^{ab}u_{A}\underline{\nabla_{c}u^{e}}\\
\nonumber&&-u^{c}(\gamma_{A}^{e}-u_{A}u^{e})\nabla_{c}(u_{b}\gamma_{ae}t_{\mathcal{S}}^{ab})-t_{\mathcal{S}}^{cd}h_{A}^{e}(u_{e}\nabla_{d}u_{c}+u_{c}\nabla_{d}u_{e})\\
\nonumber&=&(\gamma_{aA}t_{\mathcal{S}}^{ad}\gamma_{d}^{b})_{||b}-\underline{\gamma_{b}^{c}}\gamma_{A}^{e}\underline{u^{b}}\nabla_{c}(\gamma_{ae}t_{\mathcal{S}}^{ad}u_{d})-\gamma_{b}^{c}\gamma_{A}^{e}\gamma_{ae}t_{\mathcal{S}}^{ad}u_{d}\nabla_{c}u^{b}\\
\nonumber&&\underline{\underline{+t_{\mathcal{S}e}^{c}u_{A}\nabla_{c}u^{e}}}-u^{c}\gamma_{A}^{e}\nabla_{c}(u_{b}\gamma_{ae}t_{\mathcal{S}}^{ab})\\
\nonumber&&+u^{c}u_{A}\nabla_{c}(\underline{u^{e}}u_{b}\underline{\gamma_{ae}}t_{\mathcal{S}}^{ab})-\underline{u^{c}}u_{A}u_{b}\underline{\gamma_{ae}}t_{\mathcal{S}}^{ab}\underline{\nabla_{c}u^{e}}\\
\nonumber&&\underline{\underline{-t_{\mathcal{S}}^{cd}u_{A}\nabla_{d}u_{c}}}-t_{\mathcal{S}}^{cb}(\gamma_{b}^{d}-u_{b}\underline{u^{d}})(\gamma_{A}^{e}-u_{A}\underline{u^{e}})u_{c}\underline{\nabla_{d}u_{e}}\\
&=&(\gamma_{aA}t_{\mathcal{S}}^{ad}\gamma_{d}^{b})_{||b}-\gamma_{b}^{c}\gamma_{aA}t_{\mathcal{S}}^{ad}u_{d}\nabla_{c}u^{b}-u^{c}\gamma_{A}^{e}\nabla_{c}(u_{b}\gamma_{ae}t_{\mathcal{S}}^{ab})-t_{\mathcal{S}}^{cb}\gamma_{b}^{d}\gamma_{A}^{e}u_{c}\nabla_{d}u_{e}.
\end{eqnarray}
Taking the limit $\alpha\rightarrow0$, we can rewrite it in the form
\begin{eqnarray}
%\nonumber-\frac{1}{\alpha}T_{a}^{c}l_{c}\gamma_{A}^{a}&=&\frac{1}{\alpha}[\gamma_{AA'}\gamma_{B}^{b}(p_{\mathcal{H}}\gamma^{A'B}-2\eta_{\mathcal{H}}\sigma_{\mathcal{H}}^{A'B}-\zeta_{\mathcal{H}}\theta_{\mathcal{H}}\gamma^{A'B})]_{||b}\\
%\nonumber&&+\pi_{\mathcal{H}A}\gamma_{b}^{c}\nabla_{c}u^{b}+\gamma_{A}^{e}u^{c}\nabla_{c}\pi_{\mathcal{H}e}+\pi_{\mathcal{S}}^{d}\gamma_{A}^{e}\nabla_{d}u_{e}\\
\nonumber-\frac{1}{\alpha}T_{a}^{c}l_{c}\gamma_{A}^{a}&=&\frac{1}{\alpha}[\gamma_{AA'}\gamma_{B}^{b}(p_{\mathcal{H}}\gamma^{A'B}-2\eta_{\mathcal{H}}\sigma_{\mathcal{H}}^{A'B}-\zeta_{\mathcal{H}}\theta_{\mathcal{H}}\gamma^{A'B})]_{||b}\\
\nonumber&&+\frac{1}{\alpha}\pi_{\mathcal{H}A}\gamma_{b}^{c}\nabla_{c}(\alpha u^{b})+\gamma_{A}^{e}u^{c}\nabla_{c}\pi_{\mathcal{H}e}+\frac{1}{\alpha}\pi_{\mathcal{H}}^{d}\gamma_{A}^{e}\nabla_{d}(\alpha u_{e})\\
\nonumber&=&\frac{1}{\alpha}[\gamma_{AA'}\gamma_{B}^{b}(p_{\mathcal{H}}\gamma^{A'B}-2\eta_{\mathcal{H}}\sigma_{\mathcal{H}}^{A'B}-\zeta_{\mathcal{H}}\theta_{\mathcal{H}}\gamma^{A'B})]_{||b}\\
\nonumber&&+\frac{1}{\alpha}\pi_{\mathcal{H}A}\gamma_{b}^{c}\nabla_{c}l^{b}+\frac{1}{\alpha}\gamma_{A}^{e}l^{c}\nabla_{c}\pi_{\mathcal{H}e}+\frac{1}{\alpha}\pi_{\mathcal{H}}^{d}\gamma_{A}^{e}\nabla_{d}l_{e}\\
\nonumber&=&\frac{1}{\alpha}(p_{\mathcal{H}}\gamma_{A}^{B}-2\eta_{\mathcal{H}}\sigma_{\mathcal{H}A}^{B}-\zeta_{\mathcal{H}}\theta_{\mathcal{H}}\gamma_{A}^{B})_{||B}\\
\nonumber&&+\frac{1}{\alpha}\pi_{\mathcal{H}A}\gamma_{b}^{c}k_{c}^{b}+\frac{1}{\alpha}\gamma_{A}^{e}l^{c}\nabla_{c}\pi_{\mathcal{H}e}+\frac{1}{\alpha}\pi_{\mathcal{H}}^{d}\gamma_{A}^{e}\nabla_{e}l_{d}\\
&=&\frac{1}{\alpha}[p_{\mathcal{H}||A}-2(\eta_{\mathcal{H}}\sigma_{\mathcal{H}A}^{B})_{||B}-(\zeta_{\mathcal{H}}\theta_{\mathcal{H}})_{||A}]+\frac{1}{\alpha}\pi_{\mathcal{H}A}\gamma_{b}^{c}k_{c}^{b}+\frac{1}{\alpha}\gamma_{A}^{e}\mathcal{L}_{l}\pi_{\mathcal{H}e}.
\end{eqnarray}
In the third step, we have made use of the symmetry $k_{AB}=k_{BA}$. Finally, one can substitute \eqref{2kdec} into the above equation to get
\begin{equation}
-\frac{1}{\alpha}T_{a}^{c}l_{c}\gamma_{A}^{a}=\frac{1}{\alpha}[p_{\mathcal{H}||A}-2(\eta_{\mathcal{H}}\sigma_{\mathcal{H}A}^{B})_{||B}-(\zeta_{\mathcal{H}}\theta_{\mathcal{H}})_{||A}]+\frac{1}{\alpha}\pi_{\mathcal{H}A}\theta_{\mathcal{H}}+\frac{1}{\alpha}\gamma_{A}^{e}\mathcal{L}_{l}\pi_{\mathcal{H}e}
\end{equation}
and hence the Navier-Stokes equation
\begin{equation}
\gamma_{A}^{e}\mathcal{L}_{l}\pi_{\mathcal{H}e}+\pi_{\mathcal{H}A}\theta_{\mathcal{H}}=-p_{\mathcal{H}||A}+2(\eta_{\mathcal{H}}\sigma_{\mathcal{H}A}^{B})_{||B}+(\zeta_{\mathcal{H}}\theta_{\mathcal{H}})_{||A}-T_{a}^{c}l_{c}\gamma_{A}^{a}.
\end{equation}

\end{document}